\documentclass[conference]{IEEEtran}
\IEEEoverridecommandlockouts
\usepackage{cite}
\usepackage{amsmath,amssymb,amsfonts}
\usepackage{algorithmic}
\usepackage{graphicx}
\usepackage{textcomp}
\usepackage{xcolor}
\usepackage{hyperref}
\usepackage{multirow}
\usepackage{comment}
\usepackage{url}

\usepackage{bbold}

\ifCLASSOPTIONcompsoc
\usepackage[caption=false,font=normalsize,labelfont=sf,textfont=sf]{subfig}
\else
\usepackage[caption=false,font=footnotesize]{subfig}
\fi

\usepackage[linesnumbered,ruled]{algorithm2e}

\definecolor{violet}{rgb}{0.58, 0.0, 0.83}

\def\BibTeX{{\rm B\kern-.05em{\sc i\kern-.025em b}\kern-.08em
    T\kern-.1667em\lower.7ex\hbox{E}\kern-.125emX}}
\begin{document}

\title{Enhancing LR-FHSS Scalability Through Advanced Sequence Design and Demodulator Allocation}


\author{Diego Maldonado, Megumi Kaneko,~\IEEEmembership{Senior Member,~IEEE}, Juan A. Fraire,~\IEEEmembership{Senior Member,~IEEE}, \\ 
Alexandre Guitton, Oana Iova, Hervé Rivano

\thanks{D. Maldonado is with Univ Lyon, INSA Lyon, Inria, CITI, F-69621 Villeurbanne, France,  (diego.maldonado-munoz@inria.fr).}

\thanks{M. Kaneko is with the National Institute of Informatics (NII), Tokyo 101-8430, Japan (megkaneko@nii.ac.jp).}

\thanks{J. A. Fraire is with Univ Lyon, Inria, INSA Lyon, CITI, F-69621 Villeurbanne, France, the Argentinian Research Council (CONICET), Córdoba, Argentina, and the Saarland University, Germany. (juan.fraire@inria.fr).}

\thanks{A. Guitton is with Université Clermont-Auvergne, CNRS, Clermont-Auvergne-INP, LIMOS, 63000 Clermont-Ferrand, France. (alexandre.guitton@uca.fr).}

\thanks{O. Iova is with Univ Lyon, INSA Lyon, Inria, CITI, F-69621 Villeurbanne, France,  (oana.iova@inria.fr).}

\thanks{H. Rivano is with Univ Lyon, INSA Lyon, Inria, CITI, F-69621 Villeurbanne, France,  (herve.rivano@inria.fr).}}

\maketitle

\begin{abstract}
The accelerating growth of the Internet of Things (IoT) and its integration with Low-Earth Orbit (LEO) satellites demand efficient, reliable, and scalable communication protocols. Among these, the Long-Range Frequency Hopping Spread Spectrum (LR-FHSS) modulation, tailored for LEO satellite IoT communications, sparks keen interest. This work presents a joint approach to enhancing the scalability of LR-FHSS, addressing the demand for massive connectivity. We deepen into Frequency Hopping Sequence (FHS) mechanisms within LR-FHSS, spotlighting the potential of leveraging Wide-Gap sequences. Concurrently, we introduce two novel demodulator allocation strategies, namely, ``Early-Decode" and ``Early-Drop," to optimize the utilization of LoRa-specific gateway decoding resources. Our research further validates these findings with extensive simulations, offering a comprehensive look into the future potential of LR-FHSS scalability in IoT settings.
\end{abstract}


\section{Introduction}\label{Introduction}

\begin{figure}[]
\centerline{\includegraphics[width=\linewidth]{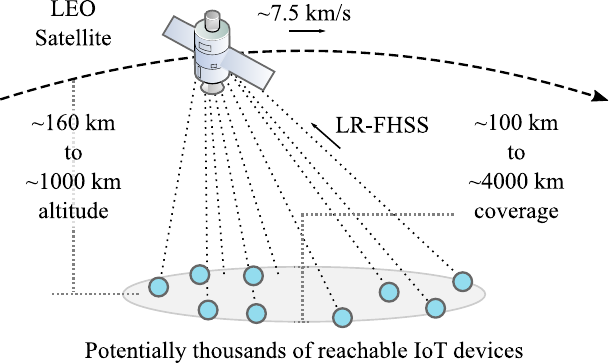}}
\caption{Direct-to-Satellite IoT Scenario.}
\label{dtsiot}
\end{figure}

Low-Earth Orbit (LEO) satellites have become instrumental in extending the coverage of various Internet of Things (IoT) applications, where terrestrial infrastructures are unsuitable due to the challenging geographical locations~\cite{fraire2022space}. These include the environmental and ecological observation of oceans, poles, expansive forests, and extensive natural parks~\cite{de2015satellite}. In these cases, LEO satellites are more cost-effective than geostationary satellites mainly because of their closer proximity to Earth, reducing transmission delays and power requirements. 
As the number of IoT devices surges, the demand for scalable and reliable communication protocols that address terrestrial and direct-to-space access (a.k.a. Direct-to-Satellite IoT~\cite{fraire2019direct}, see Fig.~\ref{dtsiot}) increases. One solution that has gained substantial attention is Long-Range (LoRa) modulation, which stands out for its low power consumption and expansive coverage. LoRa has been successfully ported to the satellite context with increasing interest from the academic and industrial communities (e.g., \href{https://www.hackster.io/news/fossasat-1-an-open-source-satellite-for-the-internet-of-things-7f31cab00ef5}{FossaSat}, \href{https://space.skyrocket.de/doc_sdat/lacunasat-3.htm}{LacunaSat}).
However, LoRa is too bandwidth-consuming, and it is challenging for a receiving device located on a satellite to decode many incoming LoRa frames. These scalability issues hinder the use of LoRa communications from Earth to satellites\cite{herreriaalonso23improving}, as the coverage area of a satellite is typically extensive.

To address the scalability issue while maintaining the high link budget required for direct-to-space communications, Semtech introduced a new modulation technique called Long-Range Frequency Hopping Spread Spectrum (LR-FHSS)~\cite{alliance2018lorawan}. The main advantage of LR-FHSS is that it can evade self-interference by using a large number of narrow-band channels and a frequency hopping technique based on a Frequency Hopping Sequence (FHS). This is done while maintaining a link budget similar to LoRa with its most robust settings. Moreover, most recent LoRa devices are compatible with the LR-FHSS modulation, streamlining LR-FHSS adoption and integration with terrestrial IoT infrastructures\footnote{As of the current writing, LR-FHSS modulation is supported in devices featuring Semtech's SX1261, SX1262, and LR1110 chipsets, and is compatible with V2.1 gateways employing the Semtech SX1301 chips~\cite{semtech2020lorawan}}.

Although the introduction of LR-FHSS is relatively recent, it has swiftly garnered attention from the research community in satellite communications. Areas of focus include performance evaluation~\cite{ullah2021analysis,boquet2021lr}, outage probabilities~\cite{maleki2022d2d,maleki2022outage}, headerless recovery~\cite{fraire2023recovering}, and transceiver design~\cite{jung2023transceiver}. These works typically assume that the FHS is random or defined by the standard. Yet, two pivotal topics that could jointly extend LR-FHSS scalability await further research and refinement: the sequence determining the frequency hop pattern, and the concurrent methodology for allocating demodulator processors to decode multiple incoming frames in parallel.

\paragraph{Sequence Design}
FHS techniques are employed to enhance the security and reliability of wireless transmissions~\cite{bao2016frequency}. These sequences quickly alternate the carrier frequency during data transmission, neutralizing interference and jamming threats~\cite{ge2009optimal}. This is achieved by dispersing the transmitted frame over an extensive frequency range, which makes it notably challenging for potential adversaries to intercept or interrupt the ongoing communication~\cite{peng2004lower}. The introduction of Wide-Gap FHS (WGFHS) is a recent advancement in this domain. This innovative sequence ensures a guaranteed minimum gap between successive frequency hops and boasts optimal attributes in terms of Hamming correlation~\cite{li2019new}. What makes WGFHS particularly noteworthy is its compliance with standard requirements set by regional regulators for maintaining gaps between adjacent frequency channels. This is precisely the case of LR-FHSS. This adheres to regulations and fortifies the anti-jamming capabilities of FHS systems~\cite{huaqing2010design, bin1997one, guan2014generation}. Despite the promising potential of this new FHS construction in the context of LR-FHSS modulation, to the authors' best knowledge, it remains an under-explored topic in existing literature.

\paragraph{Demodulator Allocation}
Existing LoRaWAN gateways, primarily powered by Semtech's SX1301 chip~\cite{semtech2017sx1301}, are equipped to concurrently demodulate several LoRa signals as long as they operate on unique channels or different Spreading Factors (SF). This SX1301 chip, a standard in most LoRaWAN gateways, houses eight distinct demodulator chips. It continuously scans for preambles across all channels and SFs, and upon detection, the integrated packet arbiter decides which of the available demodulators should demodulate the corresponding frame. The limited number of demodulators can be a potential bottleneck, impairing the network's overall performance~\cite{sorensen2019analysis,dalela2019lorawan,magrin2020study}. Addressing this limitation, authors in~\cite{guitton2020improving,guitton2022multi} proposed better algorithms to harness these demodulators in individual gateway configurations. The precise workings of the SX1301's packet arbiter for LoRa remain undocumented, and the architecture of LR-FHSS-capable chips remains unknown to the community. Indeed, LR-FHSS will likely rely on a more advanced chipset designed explicitly for demodulating FHS signals. The capabilities of these newer chipsets are still unclear, but estimates suggest that accommodating up to 1000 simultaneous demodulation might be plausible~\cite{applicationnote}. To the authors' knowledge, no demodulator allocation technique for LR-FHSS has been proposed thus far.  

\subsection{Contributions}
This paper contributes by expanding LR-FHSS scalability in two dimensions.

On the one hand, we propose to improve the FHS construction itself to enhance device-to-gateway extraction rates in view of massive connectivity. In particular, we propose leveraging Wide-Gap sequences' existing properties while making them applicable to LR-FHSS requirements. This property emerges as a superior candidate, outclassing current LR-FHSS driver implementations in specific scenarios. To the best of our knowledge, this is the first work to conduct an in-depth study of the design of LR-FHSS sequences and further exploit Wide-Gap sequences in the context of LR-FHSS modulation. In particular, we investigate The parameter design and settings of Wide-Gap sequences, such as whether they fit practical LR-FHSS requirements.


On the other hand, we introduce two pioneering strategies for demodulator allocation designed to further amplify the scalability of LR-FHSS-based networks. At the heart of these strategies is the ingenious use of redundancies inherent in the payload's convolutional encoding and decoding processes. The first, termed ``Early-Decode," frees up demodulators once they've gathered sufficient convolutional blocks for frame decoding. The second, ``Early-Drop," proactively dismisses a packet if a certain number of its blocks fail, liberating the processor for subsequent demodulation. 

The detailed contributions made in this paper are listed as follows.

\begin{enumerate}
    \item \textbf{Description of the FHS used in the driver of LR-FHSS Devices}: Based on an analysis of the source code of the driver, we describe the FHS implemented in LR-FHSS devices and show that it is based on $m$-sequences, unlike what was stated by previous research~\cite{boquet2021lr}.
    \item \textbf{Wide-Gap FHS for LR-FHSS}: By investigating the Wide-Gap FHS concept, we propose an alternative FHS construction optimized for the specifics of LR-FHSS modulation, thus enhancing its scalability.
    \item \textbf{Demodulator Optimization for LR-FHSS}: Two novel approaches (Early-Decode, Early-Drop) are introduced. Tailored to amplify the capacity of LR-FHSS demodulators, they maximize the attainable scalability.
    \item \textbf{Simulation-Driven Validation}: We evaluate the proposed techniques through an extensive simulation campaign, quantifying the practical benefits and potential implications for LR-FHSS systems. We show that our proposed FHS yields better performance than the FHS currently used in the driver of LR-FHSS devices.
\end{enumerate}

\subsection{Outline of the paper}

The remainder of the paper is organized as follows. Section~\ref{Background} presents the system model, the LR-FHSS modulation, and the state-of-the-art design of FHS. Section~\ref{Methods} delineates our proposed methods, encompassing FHS family construction and demodulator optimization. Section~\ref{NumericalResults} presents and discusses our simulation results. Section~\ref{Conclusions} draws Conclusions and avenues for future research.

\section{LR-FHSS Modulation and FHS Overview}\label{Background}

\subsection{LR-FHSS Modulation}

LR-FHSS is a new modulation technique that revolves around rapid frequency hopping within a single frame using a large number of narrow-band channels. 
Primarily tailored for uplink communications, LR-FHSS proves particularly efficient when a large number of energy-limited devices transmit simultaneously to LEO satellites, thanks to the large number of narrow-band channels LR-FHSS uses. 

\begin{figure*}
\centerline{\includegraphics[width=0.9\linewidth]{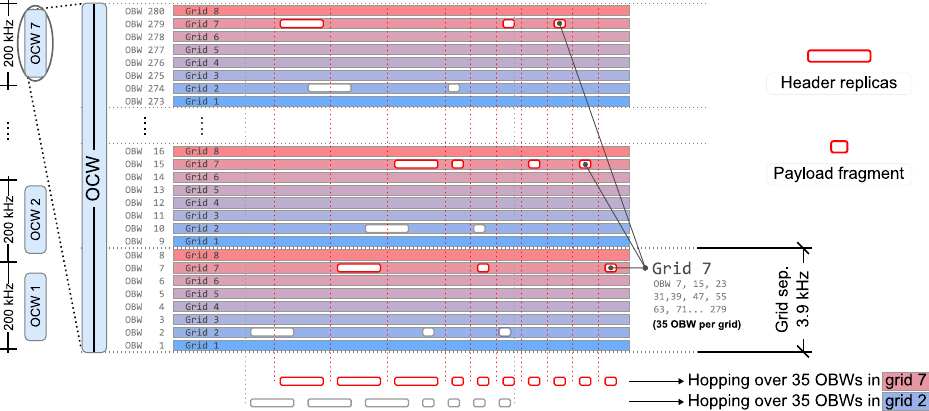}}
\caption{Division into OCWs, OBWs, and grids for 1 OCW in the European 868 MHz frequency (simplification from~\cite{fraire2023recovering})}
\label{lrfhss}
\end{figure*}

LR-FHSS segments the frequency band into a channels called Occupied Channel Widths (OCWs). Each OCW is further partitioned into multiple narrow-band channels called Occupied Bandwidths (OBWs)~\cite{boquet2021lr}. As illustrated in Fig.~\ref{lrfhss}, an individual OBW acts as a physical sub-carrier with a bandwidth set at 488 Hz. Within each OBW, LR-FHSS utilizes a Gaussian Minimum-Shift Keying (GMSK) modulation, prized for its exceptional spectral and energy efficiency~\cite{turletti1996gmsk}. To align with regional regulations stipulating the minimum spacing between sub-carriers in the FHS, OBWs are systematically arranged into grids. Grids comprise non-sequential OBWs spaced apart by a mandated minimum distance, whose value depends on regional regulatory standards (e.g., 3.9~kHz in the EU). For instance, in the EU, Grid 1 contains OBWs 1, 9, 17, $\ldots$, while Grid 2 contains OBWs 2, 10, 18, $\ldots$. These grids play a pivotal role in LR-FHSS: a hopping sequence within any given frame is restricted to using OBWs within the same grid. A summary of the LR-FHSS modulation parameters for chosen channels, including Data Rate (DR) modes and Frequency Hopping (FH) channels, is given in Table~\ref{loraparameters}. Details on the OCW channel bandwidth, OBW channel count, and the number of grids specified for LR-FHSS are provided in~\cite{applicationnote}, from which they are concisely laid out in Table 1. Note that the large number of OBW channels (between 280 and 3120) is the core of the scalability of LR-FHSS.

\begin{table}
\caption{LR-FHSS Modulation Parameters~\cite{Sornin2021-ue}}
\begin{center}
\vspace{-0.5cm}
\begin{tabular}{|c|c|c|c|c|c|c|}
\hline
Region & \multicolumn{4}{|c|}{EU863-870 MHz} & \multicolumn{2}{|c|}{US902-928} \\
\hline
LoRaWAN DR & DR8 & DR9 & DR10 & DR11 & DR5 & DR6 \\
\hline
OCW [kHz] & \multicolumn{2}{|c|}{136.70} & \multicolumn{2}{|c|}{335.94} & \multicolumn{2}{|c|}{1523.4} \\
\hline
OBW [Hz] & \multicolumn{6}{|c|}{488} \\
\hline
Grid spacing [kHz] & \multicolumn{4}{|c|}{3.9} & \multicolumn{2}{|c|}{25.4} \\
\hline
Number of grids & \multicolumn{4}{|c|}{8} & \multicolumn{2}{|c|}{52} \\
\hline
OBWs per grid & \multicolumn{2}{|c|}{35} & \multicolumn{2}{|c|}{86} & \multicolumn{2}{|c|}{60}\\
\hline
OBW channels & \multicolumn{2}{|c|}{280 (8x35)} & \multicolumn{2}{|c|}{688 (8x86)} & \multicolumn{2}{|c|}{3120 (52x60)} \\
\hline
Coding Rate & 1/3 & 2/3 & 1/3 & 2/3 & 1/3 & 2/3 \\
\hline
Header Replicas & 3 & 2 & 3 & 2 & 3 & 2 \\
\hline
Max. payload [bytes] & 58 & 123 & 58 & 123 & 58 & 133 \\
\hline
Max. $N_F$ & 31 & 32 & 31 & 32 & 31 & 34 \\
\hline
\end{tabular}
\label{loraparameters}
\end{center}
\end{table}

Frame transmission initiates with multiple header copies, called replicas, each one lasting 233 ms in airtime. Depending on the chosen data rate scheme, the header is replicated two or three times to ensure reliable reception. Note that the reception of a single replica is needed to retrieve the header. Then, replicas are followed by the segmented payload, where each segment, called a fragment, lasts 102.4 ms in airtime. These replicas and fragments are relayed over varying OBWs within a single grid, according to the FHS. Every header replica conveys the index of the chosen FHS. The multiplicity of sequences, which is typically 512, combined with the large number of OBWs, reduce the possibility of multiple frames adopting an identical hopping sequence. Payload encoding is designed to keep data integrity intact even with substantial fragment loss~\cite{boquet2021lr}. The LoRaWAN standard prescribes two LR-FHSS configurations:
\begin{itemize}
    \item Robust Configuration: Headers are transmitted three times, and the payload is encoded at CR = 1/3 (denoted as CR1 in this work), meaning that successful frame recovery demands a minimum of 33\% fragment decoding.
    \item Fast Configuration: Headers are transmitted twice, and the payload is encoded at CR = 2/3 (denoted as CR2 in this work), meaning that successful frame recovery demands a minimum of 67\% fragment decoding.
\end{itemize}

A crucial element impacting the scalability of LR-FHSS is the FHS, which fundamentally determines how LR-FHSS frames with different sequences overlap and, subsequently, the extraction rate of colliding LR-FHSS frames. This relationship encompasses the intricate interplay between LR-FHSS’s error correction mechanisms and the mandatory frequency shifts dictated by grid definitions. However, to our knowledge, there is a lack of studies on the design of appropriate FHS techniques for LR-FHSS, such as those in the Wide-Gap FHS (WGFHS) family.

%

\subsection{Frequency Hopping Sequences}

The FHS defines the sequence and timing of frequency shifts during a transmission. Often, these sequences are generated from mathematical algorithms or pseudo-random number generators. The key feature of these sequences is their unpredictability and non-repetitive nature, ensuring that potential adversaries find it challenging to intercept or jam the transmission~\cite{Torrieri2018}. Moreover, FHS-based modulations exhibit significant resilience to intense interference scenarios, typical in satellite communications, due to the expansive areas they cover on the Earth's surface.

In the following, the set of accessible frequency channels in an FHSS system is denoted $\mathcal{F}=\{f_0,f_1,\cdots,f_{\ell-1}\}$. A FHS $X$ is a periodic sequence $\{x_i, i=0\dots L\}$ of frequencies within $\mathcal{F}$. In the example of Fig.~\ref{lrfhss}, $\ell = 280$, and the period is $L = 35$.

The smallest gap or distance \(e\) between successive frequencies of $X$ is delineated as:
\begin{equation}
    e = \min_{0\leq i\leq L-2} \{|x_{i+1}-x_i|, |x_{L-1}-x_0|\}.
    \label{gap}
\end{equation}
In situations where \(e > 0\), there are no two successive frequencies that are similar, and sequence \(X\) is classified as a wide-gap FHS (WGFHS)~\cite{li2019new,li2021constructions}. 
For LR-FHSS, the concept of grids is a means to achieve such a wide gap property. 

\subsection{Hamming Correlation Metrics}\label{Metrics}

Hamming Correlation is a metric of the similarity or correlation between two FHSs~\cite{sarwate1994reed}. In our setting, it quantifies the number of collisions between two periodic sequences potentially shifted in time. 
Let \(X = \{x_i\}\) and \(Y = \{y_j\}\) be two FHS, their Hamming Correlation at a time offset \(\tau\) is:
\begin{equation}
    H(X,Y;\tau) = \sum_{i=0}^{L-1} h(x_i,y_{i+\tau\pmod L}), 
    \label{hc}
\end{equation}
where \(h = \mathbb{1}_{\{f_i=f_j\}}\).



For convenience, we recall two other basic metrics:
\begin{itemize}
    \item \textbf{Auto-Correlation (AC):} This represents the correlation of a sequence with itself, i.e., \(X = Y\). The shift value \(\tau = 0\) is not interesting for AC since it would lead to a perfect match. One can define the maximum and average AC across all possible non-zero shifts:
    \begin{eqnarray}
        H_{\mathrm{max}}(X) &=& \max_{1\leq \tau \leq L-1} H(X,X;\tau),\\
        H_{\mathrm{avg}}(X) &=& \frac{1}{L-1} \sum_{\substack{X\in S,\\ 1\leq \tau \leq L-1}} H(X,X;\tau).
        \label{maxhac}
    \end{eqnarray}
    \item \textbf{Cross-Correlation (CC):} This quantifies the similarity between two different sequences, i.e., \(X \neq Y\). One can define the maximum and average CC across all possible shifts, including zero:
    \begin{eqnarray}
        H_{\mathrm{max}}(X,Y) &=& \max_{0\leq \tau \leq L-1} H(X,Y;\tau),\\
    H_{\mathrm{avg}}(X,Y) &=& \frac{1}{L} \sum_{\substack{X,Y\in S, X\neq Y,\\0\leq \tau \leq L-1}} H(X,Y;\tau).
        \label{maxhcc}
        \end{eqnarray}
\end{itemize}

These metrics are naturally extended to a family $\mathcal S$ of FHS by averaging them, denoted as $\bar{H}_{\mathrm{max}}(\mathcal{S})$ or $\bar{H}_{\mathrm{max}}(\mathcal{S},\mathcal{S})$.

Together, these metrics—both maximum and average—offer a comprehensive understanding of the relationships and characteristics of sequences within the set \(\mathcal{S}\), crucial for optimizing and ensuring the reliability of FHSS communication systems.

\subsubsection{Desired Hamming Correlation Properties}\label{Optimality}

In practice, a family of FHS should have minimal AC and CC. 

Lempel and Greenberger~\cite{lempel1974families} derived lower bounds to the auto- and cross-correlation of sequences:
\begin{eqnarray}
      H_{\mathrm{max}}(X) \geq \biggl \lceil \frac{(L-\epsilon)(L+\epsilon-\ell)}{\ell(L-1)} \biggr \rceil,\quad \epsilon = L\mod \ell,\\
    \max \{ H_{\mathrm{max}}(X), H_{\mathrm{max}}(Y), H_{\mathrm{max}}(X,Y) \} \geq p^{n-k}.
    \label{lemgreenboundfamily}
\end{eqnarray}

Such findings opened the door to the concept of optimal FHS families. An FHS family \(\mathcal{S}\) achieves optimality if every distinct pair of sequences satisfies equality in Eq. (\ref{lemgreenboundfamily}). Lempel and Greenberger introduced \(m\)-sequences, a specific type of pseudo-random binary sequence. They proposed FHS construction techniques based on \(m\)-sequences that optimize the Hamming Correlation for generating optimal FHS families.

In a subsequent study by Li et al.~\cite{li2019new}, these bounds were extended to accommodate FHSs with an additional \textit{wide-gap} property, leading to a new threshold defined as
\begin{equation}
    H_{\mathrm{max}}(X) \geq \biggl \lceil \frac{(L-\epsilon)(L+\epsilon-\ell)}{\ell(L-3)} \biggr \rceil.
    \label{li-fanbound}
\end{equation}
$X$ is said to be an optimal WGFHS if it achieves equality in Eq. (\ref{li-fanbound}). Reference \cite{li2019new} further extends this bound for the FHS families of   ~\cite{peng2004lower}.

Next, we describe the benchmark methods for constructing FHSs: the general $m$-sequences, a WGFHS from \cite{li2021constructions}, and one specific to LR-FHSS.

\subsection{Construction Techniques}\label{FHSconstructions}

\subsubsection{Constructing Maximum Length Sequences (m-sequences)}\label{Msequence}

At the heart of many sequence generation techniques lies the Linear Feedback Shift Register (LFSR), a mechanism where the input is derived from a linear combination of prior states.
The essential characteristics of an LFSR include:
\textit{i}) \textit{Generator Polynomial}, dictating the recurrence relation for the sequence generation, and \textit{ii}) \textit{Initial State Vector}, often termed the state vector or seed, is an \(m\)-bit binary integer that initializes the sequence generation process.

\begin{figure}[h]
\centerline{\includegraphics[width=0.9\linewidth]{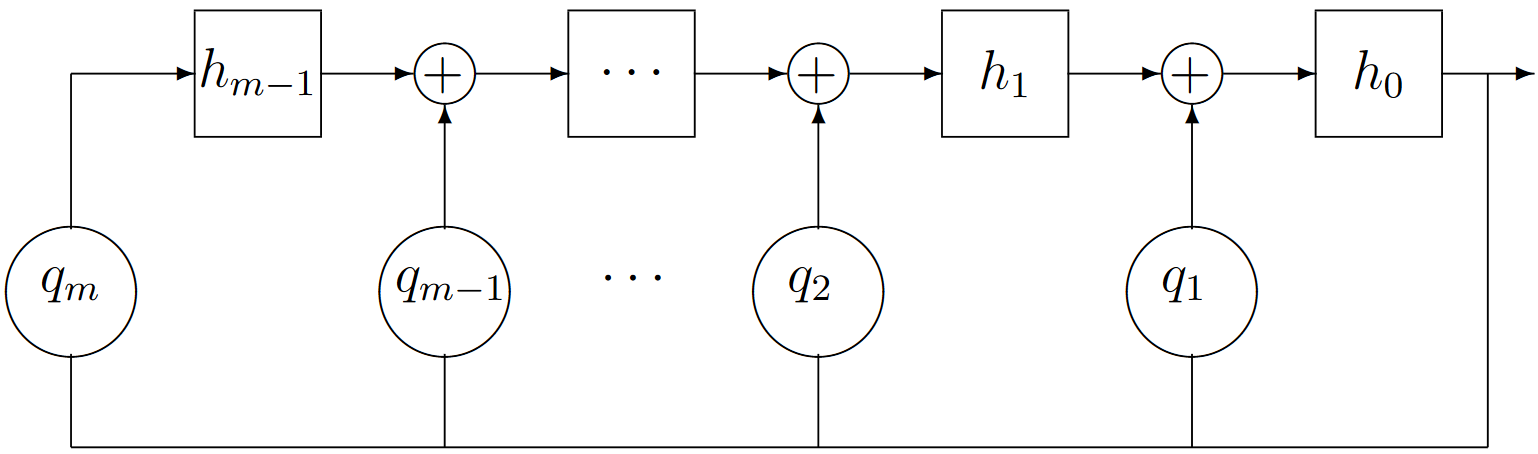}}
\caption{Galois Linear Feedback Shift Register}
\label{galois}
\end{figure}

Two primary configurations of LFSRs are recognized: the \textit{Fibonacci} and the \textit{Galois}, with references for their implementations given in~\cite{goresky2012algebraic}. For illustrative purposes, consider Fig.~\ref{galois}. It showcases a typical Galois LFSR setup with the following:
\textit{i}) \(m\) registers;
\textit{ii}) an initial state, denoted as \(h^0\), is represented in the figure as \(h\); and
\textit{iii}) a generator polynomial, \(q\), which carries binary coefficients enumerated as \(\{q_m, \cdots, q_1\}\).

The progression from one state to the next, say from \(h^0\) to \(h^1\), is mathematically represented by:
\begin{equation}
    h_i^1 = h_{i+1}^0 + q_{i+1}h_0^0 , \hspace{3mm} 0\leq i < m-1,
    \label{lfsr}
\end{equation}
\begin{equation}
    h_{m-1}^1 = q_m h_0^0 = h_0^0.
    \label{lfsrm}
\end{equation}
It is a standard practice to ensure \(q_m \neq 0\) as outlined in Eq. (\ref{lfsrm}). Using the formula in Eq. (\ref{lfsr}), the subsequent state, \(h^2\), derives its value from \(h^1\).

Now, focusing on the output, the sequence, denoted as \(s\), emerges as a pseudo-random binary sequence and is expressed as \(s=\{h_0^0, h_0^1, h_0^2, \cdots\}\). Given that the register has a capped number of potential states (precisely \(2^m\)), it's inevitable to circle back and repeat states. Yet, with a \textit{primitive polynomial} governing the LFSR, the resultant sequence, \(s\), mimics randomness and boasts an extensive cycle length. Such sequences earn the title \textit{Maximum Length Sequence} or simply \(m\)-sequence. Notably, their period stands at \(2^m-1\)~\cite{Torrieri2018}.

\subsubsection{Constructing WGFHS with Li-Fan's Approach}\label{Li-Fan}

In \cite{li2021constructions}, two 
methodologies for constructing WGFHS are presented. 

The first 
yields an optimal WGFHS with a period \(L=2\ell\). The fundamental parameters are 
\(\ell\) and \(d\), with the constraints \(1<d<\ell/2\) and $\ell$ being prime to $d$ and $d+1$.
Two sequential structures, \(s=\{s_i\}_{i=0}^{i=\ell-1}\) and \(t=\{t_i\}_{i=0}^{i=\ell-1}\), are crafted:
\begin{align*} 
s_i &= (i\cdot d)_{\ell}, \hspace{3mm} 0 \leq i < \ell, \\ 
t_i &= (i\cdot (d+1)+1)_{\ell}, \hspace{3mm} 0 \leq i < \ell.
\end{align*}
In these equations, \((\cdot)_{\ell}\) represents the smallest nonnegative residue under the modulus \(\ell\). Now, by concatenating sequences \(s\) and \(t\), a new sequence \(\mathbf{S}\) emerges with a length \(L=2\ell\). This sequence, \(\mathbf{S}\), stands out as an optimal WGFHS 
with 
\(H_{\mathrm{max}}(X)=2\) and a minimal gap of \(e=d-1\).

The second construction methodology 
differs in its outcome. It produces an optimal WGFHS with a longer period, \(L=3\ell\), over the domain \(\mathcal{F}\) with the same size \(\ell\). The sequence's defining features include \(H_{\mathrm{max}}(X)=3\) and a minimum gap of \(e=d-1\).

\subsubsection{Constructing LR-FHSS Sequences with Semtech's Driver}\label{lrfhss_sequences}

\begin{figure}[h]
\centerline{\includegraphics[width=0.9\linewidth]{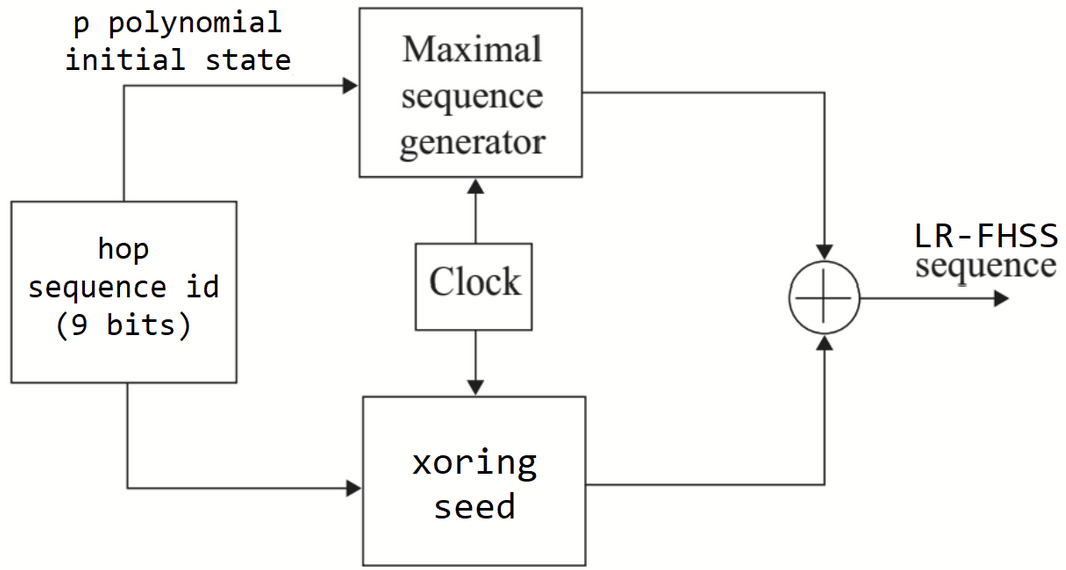}}
\caption{LR-FHSS Device's FHS Generation Scheme Using the Driver}
\label{driver}
\end{figure}

The construction technique for FHSS sequences tailored for LR-FHSS devices, as found in an official LoRa repository on Github\footnote{SX126X radio driver: \url{https://github.com/Lora-net/sx126x_driver}}, diverges from the method detailed in~\cite{boquet2021lr}. Rather than relying on a hash function, it harnesses the potential of $m$-sequences. This methodology is illustrated in Fig.~\ref{driver}.

Table~\ref{lrfhssconstructorparameters} provides an exhaustive breakdown of the LR-FHSS sequence generator's configurations assumed in the driver. Five distinct configurations arise based on the number of grids they support. Table 1 from \cite{applicationnote} lists the available LR-FHSS Channel Bandwidths, indicating the number of grids each channel supports. Several grids between 10 and 47 fall in case 1 from Table~\ref{lrfhssconstructorparameters}, 60 and 62 grids correspond to cases 2, 86 and 99 grids to case 3, 185 and 198 grids to case 4, and 390 and 403 grids to case 5. It is worth noting that for cases 1 and 2, a 6-register LFSR is employed. Yet, with this configuration, only six primitive polynomials exist, meaning the FHS families produced for LR-FHSS devices in these scenarios are limited to 384 sequences.

\begin{table}
\caption{LR-FHSS FHS Construction Parameters}
\begin{center}
\begin{tabular}{|c|c|c|c|c|c|}
\hline
case & 1 & 2 & 3 & 4 & 5 \\
\hline
initial state & 6 & 56 & \multicolumn{3}{|c|}{6} \\
\hline
LFSR size & \multicolumn{2}{|c|}{6} & 7 & 8 & 9 \\
\hline
polynomial bits & \multicolumn{2}{|c|}{3} & 2 & 1 & 0 \\
\hline
polynomial set & \multicolumn{2}{|c|}{\begin{tabular}[x]{@{}c@{}}\{33,45,48,\\ 51,54,57\}\end{tabular}} & \begin{tabular}[x]{@{}c@{}}\{65,68,\\ 71,72\}\end{tabular} & \{142,149\} & \{264\} \\
\hline
seed bits & \multicolumn{2}{|c|}{6} & 7 & 8 & 9 \\
\hline
total seeds & \multicolumn{2}{|c|}{64} & 128 & 256 & 512 \\
\hline
FHS family size & \multicolumn{2}{|c|}{384} & \multicolumn{3}{|c|}{512} \\
\hline
\end{tabular}
\label{lrfhssconstructorparameters}
\end{center}
\end{table}

The generator block for the $m$-sequence, positioned at the top of the figure, signifies an LFSR comprising $m$ registers. The primitive polynomial, instrumental in the $m$-sequence's creation, is encoded within the $x$ most significant bits of the 9-bit Hopping Sequence field conveyed in the Header, where \(x\) can take values from the set \(\{0,1,2,3\}\). The modulation parameters inform the determination of the initial state.
At the bottom, a constant \textit{xoring} seed is coded into the remaining \(y=9-x\) least significant bits of the 9-bit Hopping Sequence field. The LR-FHSS sequence is derived by performing a bit-wise exclusive OR (often referred to as \textit{xoring}, synonymous with modulo two addition) between each component of the generated $m$-sequence and the xoring seed. The sequence's length is bounded by the necessary hops to encapsulate the payload data and header replicas.
Furthermore, if the resulting values after the operation surpass the count of supported grids in a given scenario, they are omitted. 
This selective process trims down the LR-FHSS sequence's length from the original $m$-sequence length (i.e., \(2^m-1\)) to the exact count of the grids.

As a result, the \textit{driver} family is composed of the $M=384$  possible sequences that can be generated by the LR-FHSS Driver method for data rates DR8 and DR9 (case 1 from Table \ref{lrfhssconstructorparameters}), all of length $L=31$.


\section{Proposed Methods for LR-FHSS}\label{Methods}

This section describes our two main contributions to push LR-FHSS scalability limits. First, it introduces sequences crafted by harnessing the WGFHS principle delineated in Section~\ref{FHSconstructions}. These sequences are tailored to align with the stipulations of the LR-FHSS regulations. Second, we propose various strategies to enhance the LoRa demodulators' decoding proficiency in the LEO satellite scenario.

\subsection{FHS Families for LR-FHSS}\label{fhs_families}

This section delves into constructing FHS families tailored for the LR-FHSS protocol. The families constructed can be categorized based on the design principles derived from Lempel-Greenberger (see Section~\ref{Msequence}) and Li-Fan (see Section~\ref{Li-Fan}).

\subsubsection{Lempel-Greenberger FHS Families}

The Lempel-Greenberger FHS construction involves selecting parameters \(\{p, k, n\}\) and a primitive polynomial for an LFSR of a given order, generating an m-sequence that serves as the base sequence to build optimal sequences. This method enables the generation of \(M=p^k\) sequences of length \(q=p^n-1\).

\begin{itemize}
    \item {\textbf{lem-green} Family}:
    \begin{itemize}
        \item Parameters: \(p=2, k=5, n=5\),
        \item Primitive Polynomial: One of the 6 primitive polynomials (0x12) available for LFSR of order 5,
        \item Outcome: 32 sequences of length 31.
    \end{itemize}

    \item {\textbf{lem-green-2x} Family}:
    \begin{itemize}
        \item Composition: Merge of the \textit{lem-green} family and a second family using the same parameters but with a different primitive polynomial (0x14),
        \item Outcome: 64 sequences of length 31.
    \end{itemize}
\end{itemize}

\subsubsection{Li-Fan FHS Families}

The Li-Fan FHS construction provides families based on the methods of period \(2\ell\) or \(3\ell\). Depending on the chosen parameters, families might need further splitting or merging to attain desired frequency values and lengths. For all of the following Li-Fan, the frequency values above 280 are discarded (DR8 and DR9 from Table \ref{loraparameters}), and the minimum gap is adjusted to 8.

\begin{itemize}
    \item {\textbf{li-fan-2l} Family}:
    \begin{itemize}
        \item Method: \(2\ell\)-method,
        \item Parameters: \(\ell=281, d=8\),
        \item Construction Specifics: Single FHS of length \(2\ell=562\), further split into 18 sequences of length 31.
    \end{itemize}

    \item {\textbf{li-fan-2l-4x} Family}:
    \begin{itemize}
        \item Composition: Merged family from \textit{li-fan-2l} and three other Li-Fan families with parameters \(\ell=277\), \(\ell=283\) and \(\ell=287\), all with \(d=8\).
    \end{itemize}
    
    \item {\textbf{li-fan-3l} Family}:
    \begin{itemize}
        \item Method: \(3\ell\) FHS construction,
        \item Parameters: Same as the \textit{li-fan-2l} family.
    \end{itemize}

    \item {\textbf{li-fan-3l-4x} Family}:
    \begin{itemize}
        \item Composition and method: Analogous to the \textit{li-fan-2l-4x} family.
    \end{itemize}
\end{itemize}

We tailored FHS families optimized for the LR-FHSS protocol by employing the above design principles and methods. These are crucial as they lay the foundation for enhancing the decoding proficiency of LoRa demodulators, a significant step forward for satellite-based communications.
 
\subsection{Early Drop and Early Decode mechanisms for LR-FHSS}\label{Demodulators}

The satellite-based LR-FHSS gateway's primary functionality can be conceptualized as follows:
\begin{itemize}
    \item \textit{Frame Reception}: Upon the LoRa gateway's reception of a transmission from a satellite, an available demodulator springs into action. Its core responsibilities encompass collision monitoring and the capture of headers and any fragmented data in the transmission.
    \item \textit{Frame Demodulation}: After completing the reception process of all fragments, the actual demodulation occurs. 
    A packet is successfully decoded if at least one header replica and a minimum number of necessary fragments are free from collisions. 
    \item \textit{Demodulator Availability}: When no demodulators are on standby or accessible, the transmission is sidestepped. As a result, the corresponding packets are discarded.
\end{itemize}

This structured approach streamlines the gateway's operation, ensuring efficiency and minimal data loss.

Given the encoding structure of the payload, there's potential to decode it successfully, even if only a portion of the fragments are available. This suggests an opportunity to utilize each demodulator's capability more efficiently. Towards this end, we propose adopting the subsequent smart LoRa gateway mechanisms.

\subsubsection{Early Drop}\label{EarlyDrop}

\begin{figure}[h!]
\centerline{\includegraphics[width=\linewidth]{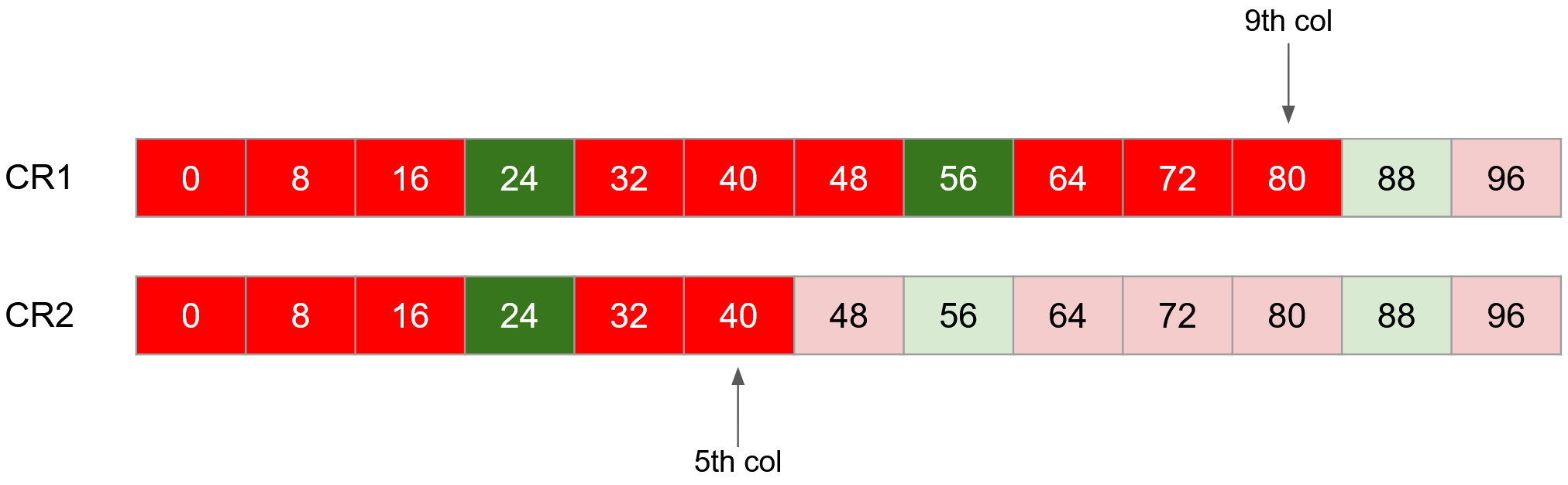}}
\vspace{-0.45cm}
\caption{Illustration of Early Drop for a 13-fragment payload}
\label{earlydrop}
\end{figure}

The Early Drop demodulation strategy is specifically designed to optimize the efficiency of the satellite gateway's demodulation resources. This mechanism comes into play when the demodulator determines that the minimum number of fragments required for successful packet decoding cannot be achieved. Under such circumstances, the demodulator preemptively terminates the reception of the packet.
To elucidate the concept, consider a payload consisting of 13 fragments. Depending on the encoding scheme:
\begin{itemize}
    \item Code Rate 1 (CR1): A minimum of 5 fragments is essential for successful decoding.
    \item Code Rate 2 (CR2): 9 fragments are required.
\end{itemize}
Figure \ref{earlydrop} delineates these configurations, where each fragment corresponds to a frequency channel. Red highlights collisions (unsuccessfully received fragments), while green signifies successfully received fragments.

In a CR1 scenario, after recognizing the 9th collision, it becomes clear that only a maximum of 4 successful fragments can be accumulated, which is inadequate for decoding. Recognizing this, the demodulator can promptly categorize the payload as 'collided' and free itself for other tasks. Analogously, for CR2, the demodulator would adopt a similar course of action after the 5th collision. The steps involved in this process are detailed in Algorithm~\ref{alg:EarlyDrop}.

\begin{algorithm}
\caption{Early Drop}\label{alg:EarlyDrop}

totalFragments $\gets$ 31

\While{\text{fragments are being received}}{
collidedFrags $\gets$ number of collided fragments

threshold $\gets$ minimum fragments for decoding

\If{collidedFrags $>$ (totalFragments-threshold)}{

drop packet due to payload collision

terminate current reception

\textbf{Return}
}
}

\end{algorithm}

\subsubsection{Early Decode}\label{EarlyDecode}
\begin{figure}
\centerline{\includegraphics[width=\linewidth]{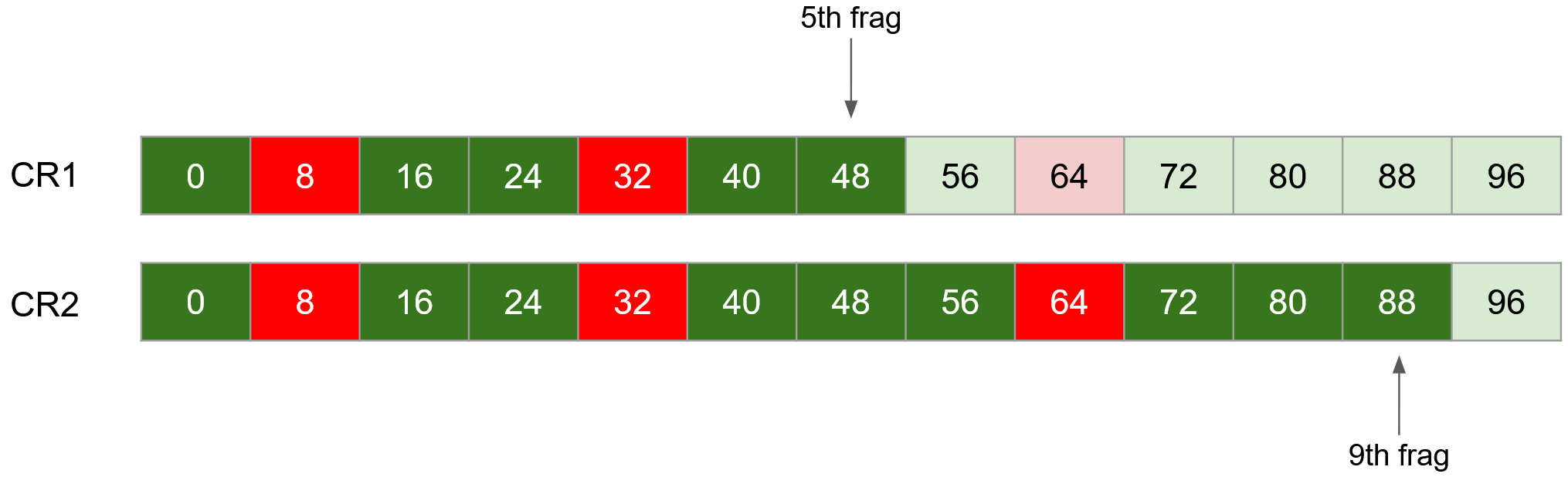}}
\vspace{-0.45cm}
\caption{Representation of Early Decode for a 13-fragment payload}
\label{earlydecode}
\end{figure}

This mechanism enables the demodulator to commence decoding once it has collected the minimum number of fragments required for successful payload decoding. This optimization lets the demodulator quickly process each packet, reducing its LoRa gateway occupancy and enhancing the overall throughput and efficiency.
Using the same fragment configurations detailed above:
\begin{itemize}
    \item Code Rate 1 (CR1): The decoding process initiates immediately after the successful reception of the 5th fragment.
    \item Code Rate 2 (CR2): Decoding begins after the successful receipt of the 9th fragment.
\end{itemize}
Fig.~\ref{earlydecode} showcases these configurations with various fragment collision statuses. Once the required fragments are received successfully, the demodulator has adequate data to decode the payload. It can then ready itself for the reception of subsequent packets. Algorithm~\ref{alg:earlydecode} offers an in-depth procedural description of the Early Decode mechanism.

\begin{algorithm}
\caption{Early Decode}\label{alg:earlydecode}

validFrags $\gets$ number of successful fragments

threshold $\gets$ minimum fragments for decoding 

\If{validFrags $\geq$ threshold}{

collidedHdrs $\gets$ number of collided headers

\If{collidedHdrs $<$ headerReplicas}{
successfully decoded frame
}

received decodable payload

terminate current reception

\textbf{Return}
}
\end{algorithm}

\section{Results and Analysis}\label{NumericalResults}

\subsection{Simulation Settings and Parameters}\label{simparams}

We constructed a simplified but representative simulation model to assess the performance implications of various FHS families in an LR-FHSS network. This model encapsulates fragments' and headers' time-slotted communication dynamics within finite frequency channels.

\subsubsection{Simulation Model Overview}

Each fragment in the simulation persists for $g$ time slots, where $g$ denotes the model's granularity. Correspondingly, headers last for $\lceil{}233g/102.4\rceil{}$ slots. Throughout the simulation duration, $N$ nodes transmit a packet of fixed length $L$, broken down into fragments. These transmissions are initiated at random time slots. The choice of an FHS for each node is randomized from the available FHS family. The gateway employed in the model is equipped with $p$ demodulators, essential for decoding the packets.

A collision is deemed to have occurred if two or more fragments or headers share at least one time slot on an identical frequency channel. Payloads receive a 'collided' designation if their collided fragments breach a certain threshold, contingent on the CR utilized for encoding. Additionally, the \textit{Header Tolerance} parameter determines the header's collision tolerance, defining the maximum collided slots permissible for a successful decode.

Our model leverages the free space channel model, which is especially fitting for LR-FHSS in satellite communications due to outer space's open and unobstructed nature. Contrary to the cluttered terrestrial environments, space provides a clear signal path, validating the model's straightforwardness as an effective means of characterizing signal interactions between satellites and terrestrial stations~\cite{vogelgesang2021uplink}. Such a channel becomes even more pertinent when factoring in the vast distances involved, which are not typically accounted for in standard urban and suburban IoT~\cite{gonzalez2021analysis}. Furthermore, frequency-selective fading is negligible, given the narrow-band channels in LR-FHSS.

Our simulator's codebase comprising the scenario configurations used in this section is publicly accessible on Github\footnote{LR-FHSS simulator: \url{https://github.com/diegomm6/lr-fhss_seq-families}}.

\subsubsection{Early Header Drop}\label{HeaderDrop}

In our extended gateway model, we introduce the ``Early Header Drop" feature to portray an LR-FHSS gateway's operation accurately. The importance of a header in packet communication cannot be overstated, as it carries the FHS information crucial for decoding the corresponding packet. Thus, if a header is not successfully decoded, the processor remains uninformed about the payload fragment sequence. In such situations, the gateway's logical course of action would be to release the demodulator, enabling it to track subsequent packets. The previous operation mode is how the LR-FHSS works in reality, so the benefit of including this feature in our simulation is to assess the potential of decoding headerless frames. This is the case when, hypothetically, the gateway receives the minimum number of fragments for a successful payload decoding.

While our research primarily focuses on assessing FHS performance within LR-FHSS networks, evaluating the ``Early Header Drop" mechanism is largely excluded from most tests in Section~\ref{NumericalResults}. Still, the study of decoding payloads without their associated headers holds considerable potential and significance in LR-FHSS research, as noted in~\cite{fraire2023recovering}.  The ``Early Header Drop" mechanism is executed after the reception of the last header and is listed in Algorithm~\ref{alg:HeaderDrop}. 

\begin{algorithm}
\caption{Early Header Drop}\label{alg:HeaderDrop}

collidedHdrs $\gets$ count of collided headers

\If{collidedHdrs $==$ headerReplicas}{
collidedPackets $\gets$ collidedPackets +1
}

terminate current reception

\textbf{Return}

\end{algorithm}

\subsubsection{Simulation Configuration}

The simulations were primed using the parameters outlined in Table~\ref{table:simparams}, following the parameters from Table~\ref{loraparameters} for DR8 and DR9. In a gateway containing a single DSP and employing a channel spacing of 200 kHz, end devices can utilize a maximum of 7 OCW channels when employing a 136.70 kHz channel bandwidth \cite{boquet2021lr}. The payload size was set to 31 fragments, the maximum allowed for DR8. The fragment granularity is six time slots. Then, the maximum sequence duration arises during CR1 encoding (factoring in 3 header repetitions), culminating in a transmission span of 228 time-slots. In contrast, CR2-encoded transmissions wrap up in 214 time slots. We chose the simulation time as a fourfold duration of the longest sequence, totaling 912 time-slots.

The ratio of fragment-to-header durations in our simulation, which stands at 14 slots / 6 slots $\approx 2.33$, is a close reflection of real-world values approximated at 233 ms / 102.4 ms $\approx 2.27$.

\begin{table}
\caption{Simulation Configuration}
\begin{center}
\vspace{-0.3cm}
\begin{tabular}{|c|c|}
\hline
Parameter & Value  \\
\hline
\hline
Number of repetitions & 10\\
\hline
Simulation time & 912 [slots]\\
\hline
OCW channels & 7\\
\hline
OBW channels & 280\\
\hline
Payload size & 31 [fragments]\\
\hline
Fragment size & 6 [slots]\\
\hline
Header size & 14 [slots]\\
\hline
\end{tabular}
\label{table:simparams}
\end{center}
\end{table}

\subsubsection{Parameters Under Study}

Table~\ref{studyparams} delineates the parameters that underwent scrutiny in our research, each having two feasible values. Both the \textit{Early Decode} and \textit{Early Drop} features can be toggled on or off individually. Nevertheless, both are activated or deactivated simultaneously in our experiments. When \textit{Early Header Drop} is turned on, the demodulator will drop the payload if no header is successfully received. For most experiments, the \textit{Header Tolerance} is set to 0, meaning that a single collided time slot will result in a collided header. The last set of experiments is set to 4 time slots.
Given that the payload fragment granularity requires six time-slots, a header will only tolerate collisions at the header reception's start, the end, or both. Our control parameter, the node count, is the transmission tally since each node transmits once within the simulation timeframe. Across different scenarios in the ensuing sections, the node (or transmission) count oscillates between 10 and 10,000. Given the results presented in section \ref{HCanalysis}, the families of interest are the driver and the Li-Fan.

\begin{table}
\caption{Study Parameters}
\begin{center}
\vspace{-0.3cm}
\begin{tabular}{|c|c|}
\hline
Parameter & Possible values\\
\hline
\hline
Coding Rate & CR1, CR2\\
\hline
FHS family & li-fan, driver\\
\hline
Number of demodulators & 100, 1000\\
\hline
Early Decode & True, False\\
\hline
Early Drop & True, False\\
\hline
Early Header Drop & True, False\\
\hline
Header tolerance in slots& 0, 4 \\
\hline
\end{tabular}
\label{studyparams}
\end{center}
\end{table}

In their work, Boquet et al.~\cite{boquet2021lr} pointed out that the FHS generation method in LR-FHSS pivots around a hash function. Drawing inspiration from this methodology, we formulated a \textit{hash} family for FHS based on the sha256 function. This curated FHS family encompasses $M=384$ families, each with a length of $L=31$.

\subsection{Preliminary Frequency Hopping Sequence Analysis}

\subsubsection{Hamming Distance Analysis}
\label{HCanalysis}

\begin{figure*}[!t]
\centering
\subfloat[Maximum Hamming Correlation]{\includegraphics[width=1\columnwidth]{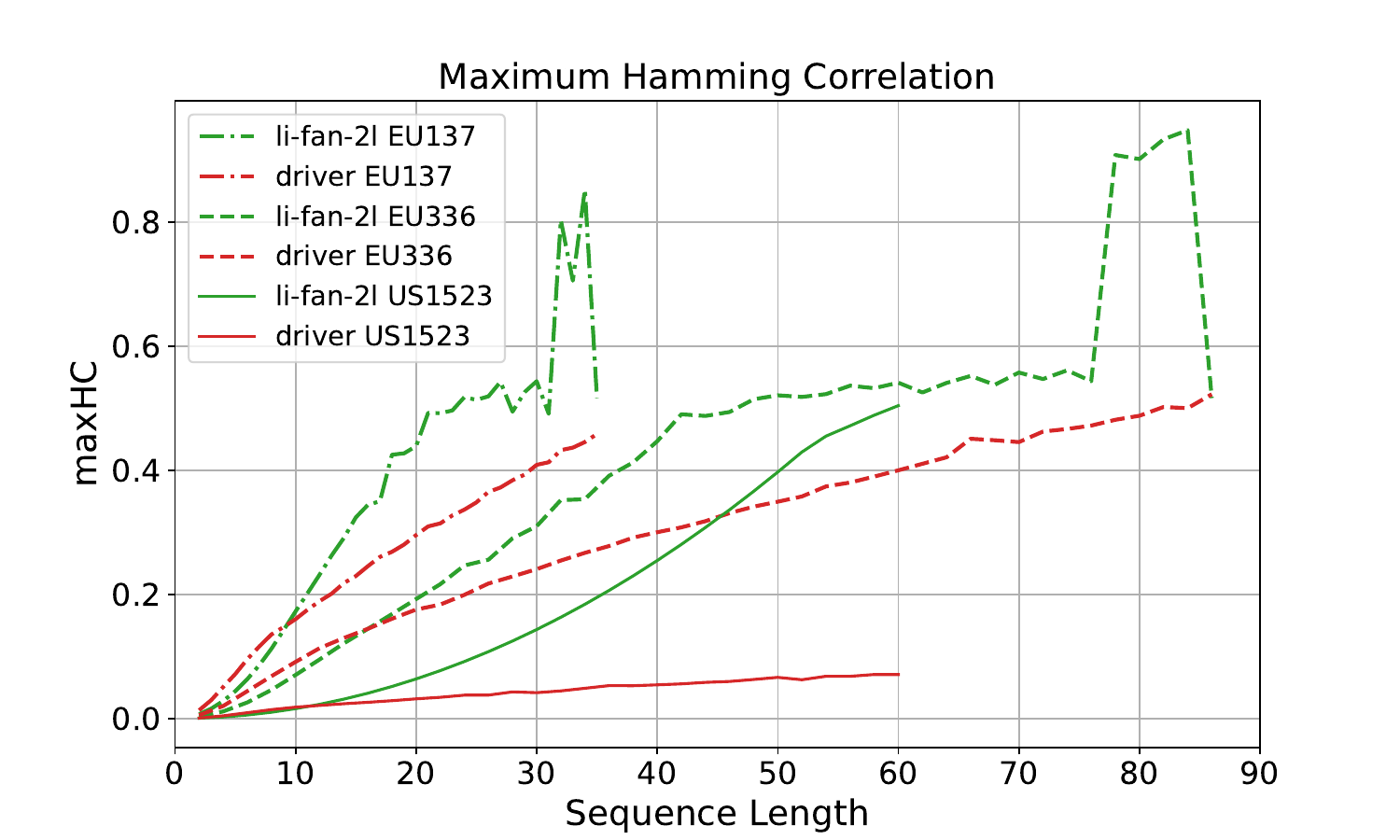}
\label{maxHC}}
\hfil
\subfloat[Average Hamming Correlation]{\includegraphics[width=1 \columnwidth]{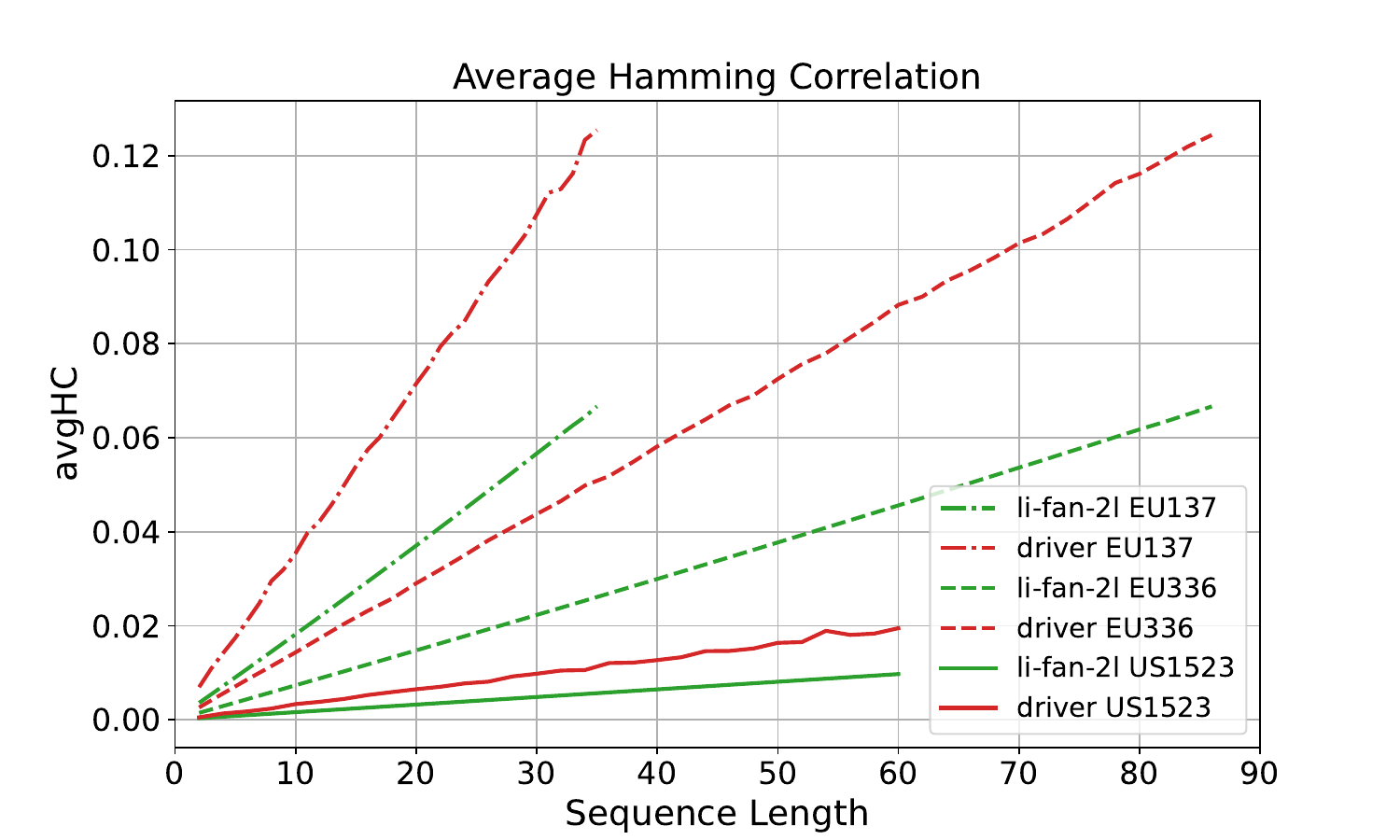}
\label{avgHC}}
\vspace{-0.1cm}
\caption{Maximum and average Hamming Correlation for different LR-FHSS FHS families (\textit{li-fan} and \textit{driver}).}
\vspace{-0.4cm}
\label{hcresults}
\end{figure*}

The summary of the evaluated FHS families can be found in Table~\ref{families}, where they are distinguished based on their size and the family's maximum Hamming Cross-Correlation $\bar{H}_{\mathrm{max}}(S,S)$.

\begin{table}
\caption{FHS family size and correlation comparison}
\begin{center}
\vspace{-0.3cm}
\begin{tabular}{|c|c|c|}
\hline
family & size & $\bar{H}_{\mathrm{max}}(S,S)$ \\
\hline
lem-green & 32 & 0.939\\
\hline
lem-green-2x & 64 & 2.108\\
\hline
li-fan-2l & 18 & 0.480\\
\hline
li-fan-2l-4x & 71 & 1.769\\
\hline
li-fan-3l & 26 & 0.704\\
\hline
li-fan-3l-4x & 107 & 1.566\\
\hline
hash & 384 & 0.407\\
\hline
driver & 384 & 0.416\\
\hline
\end{tabular}
\label{families}
\end{center}
\end{table}

The family maximum Hamming Correlation is illustrated in Figure~\ref{maxHC} for both the \textit{li-fan} and the \textit{driver} families. The correlation is assessed against increasing sequence lengths. For shorter sequences spanning up to 10 hops, the \textit{li-fan} family exhibits a reduced correlation compared to the \textit{driver} family. However, as the sequence length extends beyond this range, the \textit{driver} family takes precedence in terms of lower correlation.
It is noteworthy that even though the \textit{li-fan} FHS is designed to yield an optimal sequence, this property is guaranteed only for the singular sequence generated. This explains the observed degradation in correlation performance when this primary sequence is partitioned into more minor sequences to constitute an LR-FHSS-compatible FHS family.

Fig.~\ref{avgHC} presents the family average Hamming Correlation $\bar{H}_{\mathrm{avg}}(S,S)$. 
This correlation is demonstrated for both \textit{driver} and \textit{li-fan} families, each formulated under three distinct configuration parameters, as detailed in Table~\ref{loraparameters}. The three configurations are: 1. \textit{EU137} with a maximum FHS length of 35 hops, 2. \textit{EU336} with a maximum of 86 hops, and 3. \textit{US1523} with 60 hops.
Interestingly, across all these configurations, the \textit{li-fan} FHS families consistently outperform their \textit{driver} counterparts, showcasing superior correlation metrics.

\subsubsection{Collision Rate Analysis}

\begin{figure*}[tb]
\centerline{\includegraphics[width=1\linewidth]{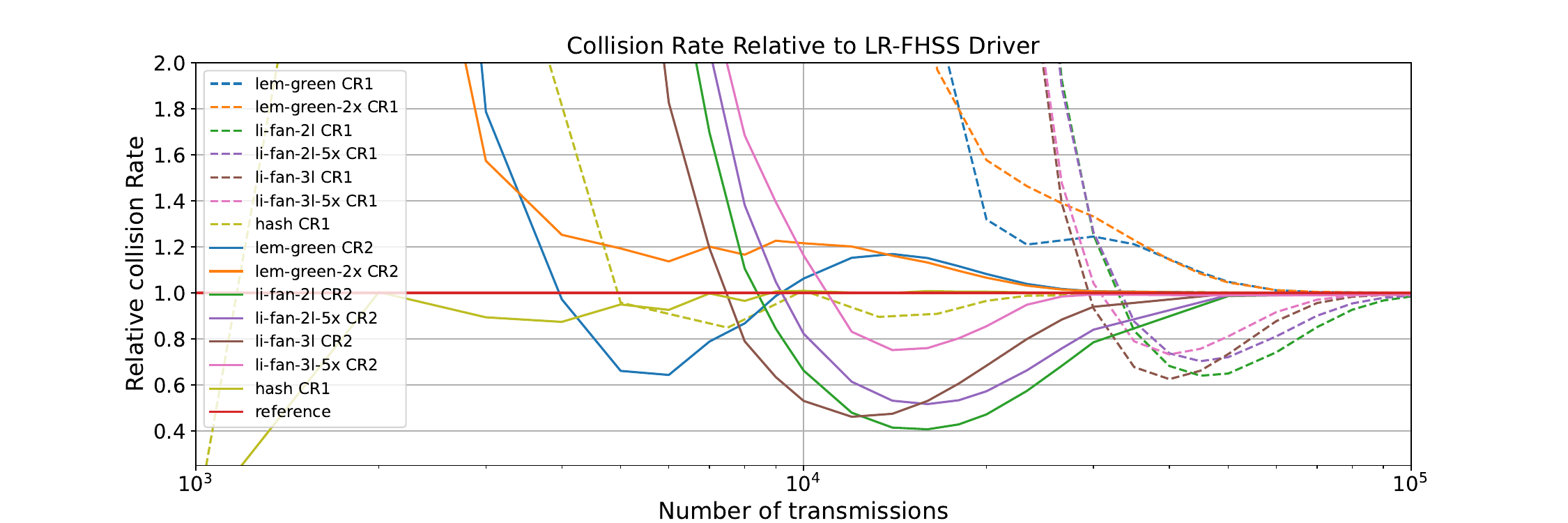}}
\vspace{-0.3cm}
\caption{Relative collision rate for different LR-FHSS FHS families, considering no granularity, no headers, and unlimited decoding capacity.}
\label{nogranularity}
\end{figure*}

For an initial assessment, we used a preliminary test to study the collision rate at the payload level for different FHS families, as detailed in Table~\ref{families}. This model is more basic than the one discussed above, encompassing fewer parameters. The collision rate is calculated as the number of collided slots from overlapping transmissions, i.e., with more than one fragment, divided by the total slots available.
This design eliminated granularity, meaning each fragment was equivalent to a single time slot. Throughout the simulation, each node was responsible for transmitting a single payload of 31 fragments devoid of headers. The total time allocated for simulation was fourfold the time required for payload transmission, summing up to 124 time-slots. Another distinguishing feature of this model was its lack of a decoding threshold for simultaneous receptions.

The core aim of this basic model was to analyze the FHS's correlation performance under conditions of multiple concurrent transmissions. Fig.~\ref{nogranularity} presents the relative collision rate, which is compared against the baseline established by the \textit{driver} family. Here, the collision rate of each family is relative to the one achieved by the \textit{driver} family for every transmission count.
The \textit{lem-green} family initially demonstrated a better performance than the \textit{driver} family up to a certain number of transmissions but later faced a decline, especially for CR2. Observations highlighted that the \textit{li-fan} families exhibited superior performance compared to the reference point after reaching a specific number of nodes. This trend was consistent for both Coding Rates.

Given the insights derived from this preliminary evaluation, we select the \textit{li-fan-2l} family as the most performing in LR-FHSS. The upcoming comprehensive simulation campaigns will further investigate the combined performance of \textit{li-fan-2l} with the proposed demodulation optimization techniques.

\subsection{Early Drop and Early Decode Analysis}

\begin{figure*}
\centering
\subfloat[CR1]{\includegraphics[width=1 \columnwidth]{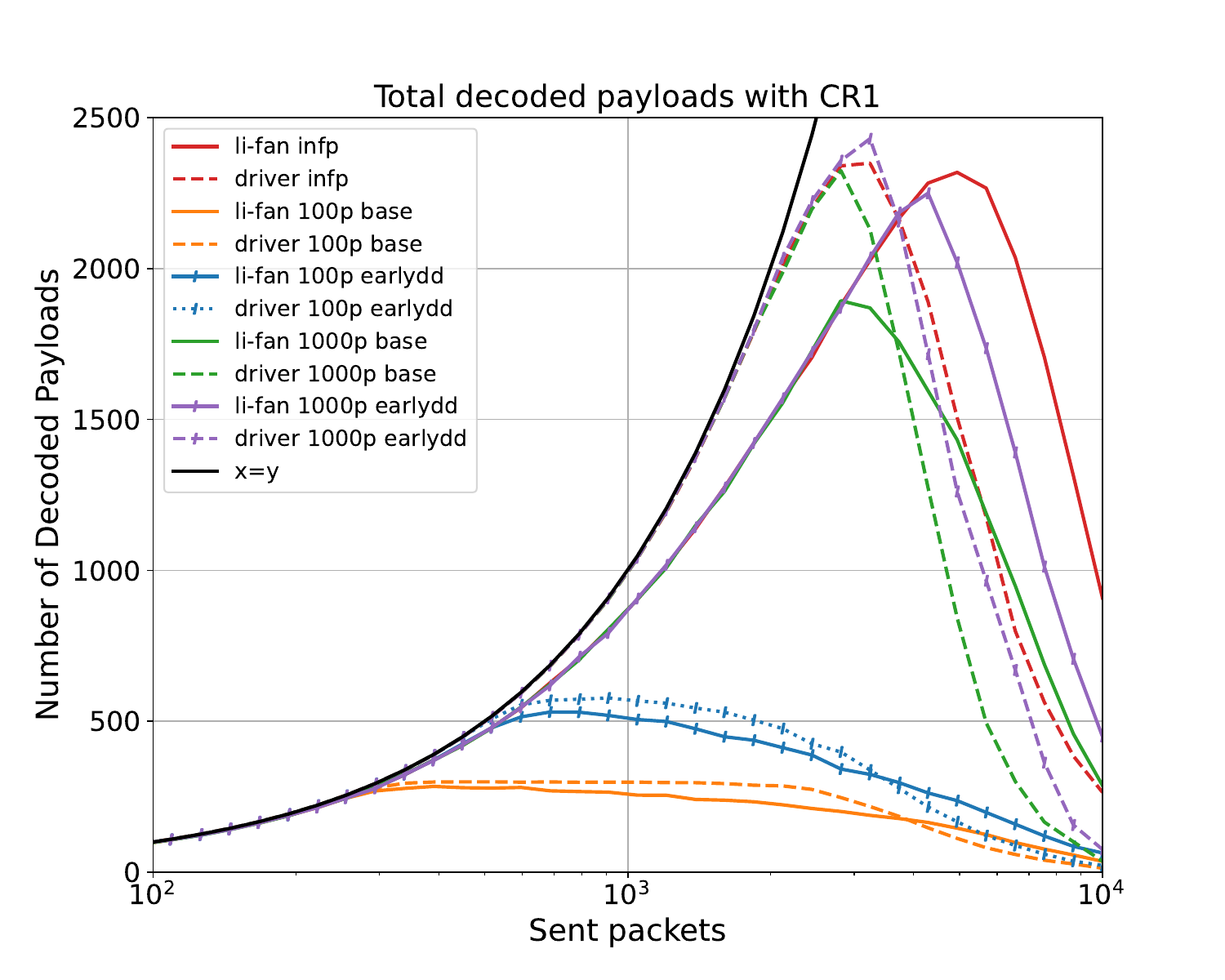}
\label{earlyddcr1}}
\hfil
\subfloat[CR2]{\includegraphics[width=1 \columnwidth]{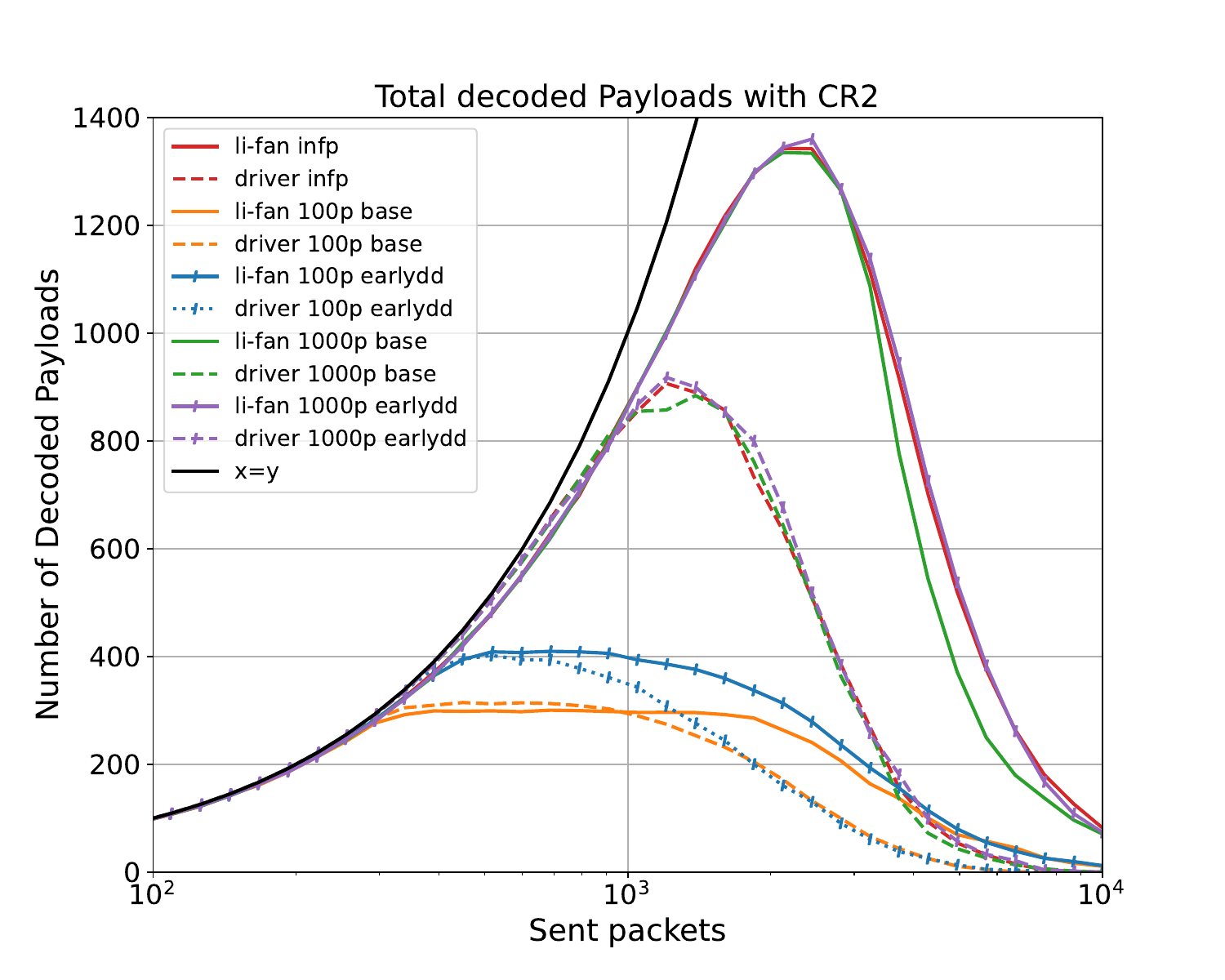}
\label{earlyddcr2}}
\vspace{-0.2cm}
\caption{Total decoded payloads with CR1 and CR2.}
\label{earlydd}
\end{figure*}

Building on our selection of the \textit{li-fan} family, our analysis proceeds with the comprehensive model detailed in Section~\ref{simparams}. Under this framework, a packet can experience one of three outcomes:
\begin{enumerate}
\item \textit{Discarded}: When no demodulators are available.
\item \textit{Successful}: A demodulator decodes the payload.
\item \textit{Collided}: The payload fails to be decoded as the minimum number of fragments is unmet.
\end{enumerate}
Although headers are not essential for payload decoding in this scenario, their presence still influences collision dynamics. We distinguish between \textit{payload} decoding, when the minimum amount of fragments is received, and the payload can be successfully decoded, and \textit{packet} decoding, when at least one header replica is received along with the payload.

Fig.~\ref{earlyddcr1} plots the cumulative decoded payloads against the total packets transmitted using CR1. The comparison is multifaceted:
\begin{itemize}
\item Families Comparison: The solid lines represent the \textit{li-fan} family, while the dashed lines depict the \textit{driver} family.
\item Decoding Mechanisms: Symbol \textit{I} marks the results derived from the \textit{Early Drop} and \textit{Early Decode} mechanisms, tagged as \textit{earlydd}, which are set against the baseline demodulator model.
\item Demodulators Count: The data segregates between setups with 100 demodulators (depicted in shades of orange and blue) and 1000 demodulators (showcased in shades of green and purple). For reference, the unlimited decoding capacity scenario is illustrated with the label \textit{infp} in red.
\end{itemize}


With the configuration of 100 demodulators (100p), activating the \textit{earlydd} mechanism brings about a significant enhancement in decoding performance for both the \textit{li-fan} and \textit{driver} families. Specifically, the number of successfully decoded payloads can see an increase of up to 100\%, effectively doubling the decoding output.
However, when the demodulator count is raised to 1000 (1000p), the benefits of the \textit{earlydd} mechanism become less pronounced. Even so, the decoding capability curve is nudged closer to that of the model with infinite decoding capacity. A notable observation is that, under this configuration, the \textit{li-fan} family witnesses an improvement of up to 50\% in successfully decoded payloads when the total transmitted packet count hovers around 4000.

When contrasting the performance of the \textit{li-fan} and \textit{driver} FHS families, the \textit{driver} FHS consistently outperforms the \textit{li-fan} up to a specific threshold of transmissions. This superiority becomes even more pronounced when operating with 1000 demodulators (1000p). Beyond this threshold, however, a transition from the \textit{driver} FHS to the \textit{li-fan} FHS has the potential to yield better network throughput.

As illustrated in Fig.~\ref{earlyddcr2} for the case using CR2, the \textit{earlydd} mechanism consistently enhances the decoding capacity for both the \textit{li-fan} and \textit{driver} FHS families. The improvement is particularly significant when utilizing 100 demodulators (100p), but it tapers off as the channel nears saturation. The benefits are less pronounced when employing 1000 demodulators (1000p).

A comparative evaluation of the \textit{li-fan} and \textit{driver} families reveals a considerable advantage for the \textit{li-fan} family. This superiority is especially evident in the 1000p setup, where the decoding capacity of the \textit{li-fan} family surpasses that of the \textit{driver} family by a remarkable 170\%. While the \textit{driver} family does exhibit marginally superior performance under lighter network loads, the disparity at higher loads suggests that the \textit{li-fan} family is the best choice when CR2 is in use.

\subsection{Packet versus Payload Analysis}

\begin{figure*}[!t]
\centering
\subfloat[100 demodulators]{\includegraphics[width=1 \columnwidth]{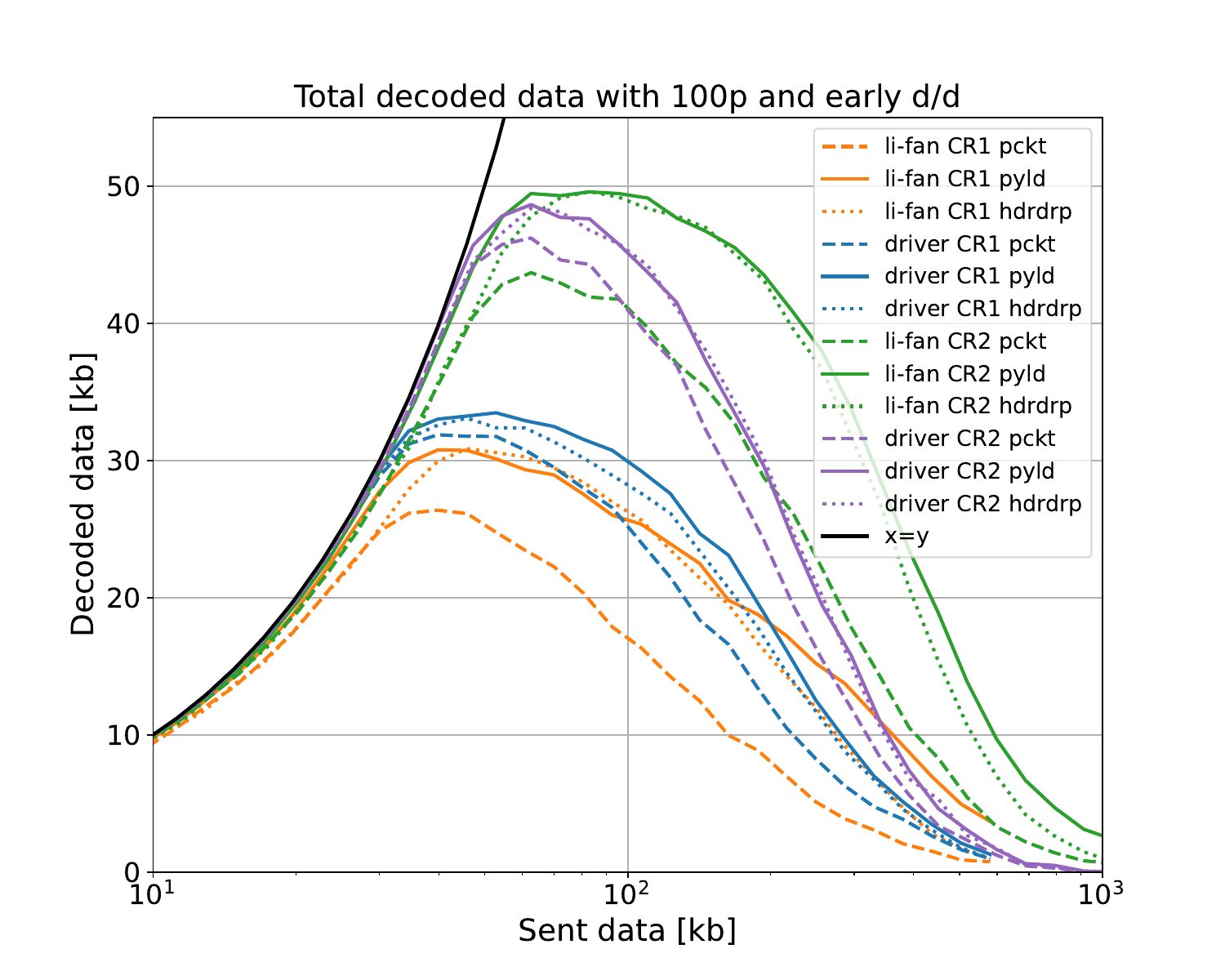}
\vspace{-0.3cm}
\label{hdrdrop100p}}
\hfil
\subfloat[1000 demodulators]{\includegraphics[width=1 \columnwidth]{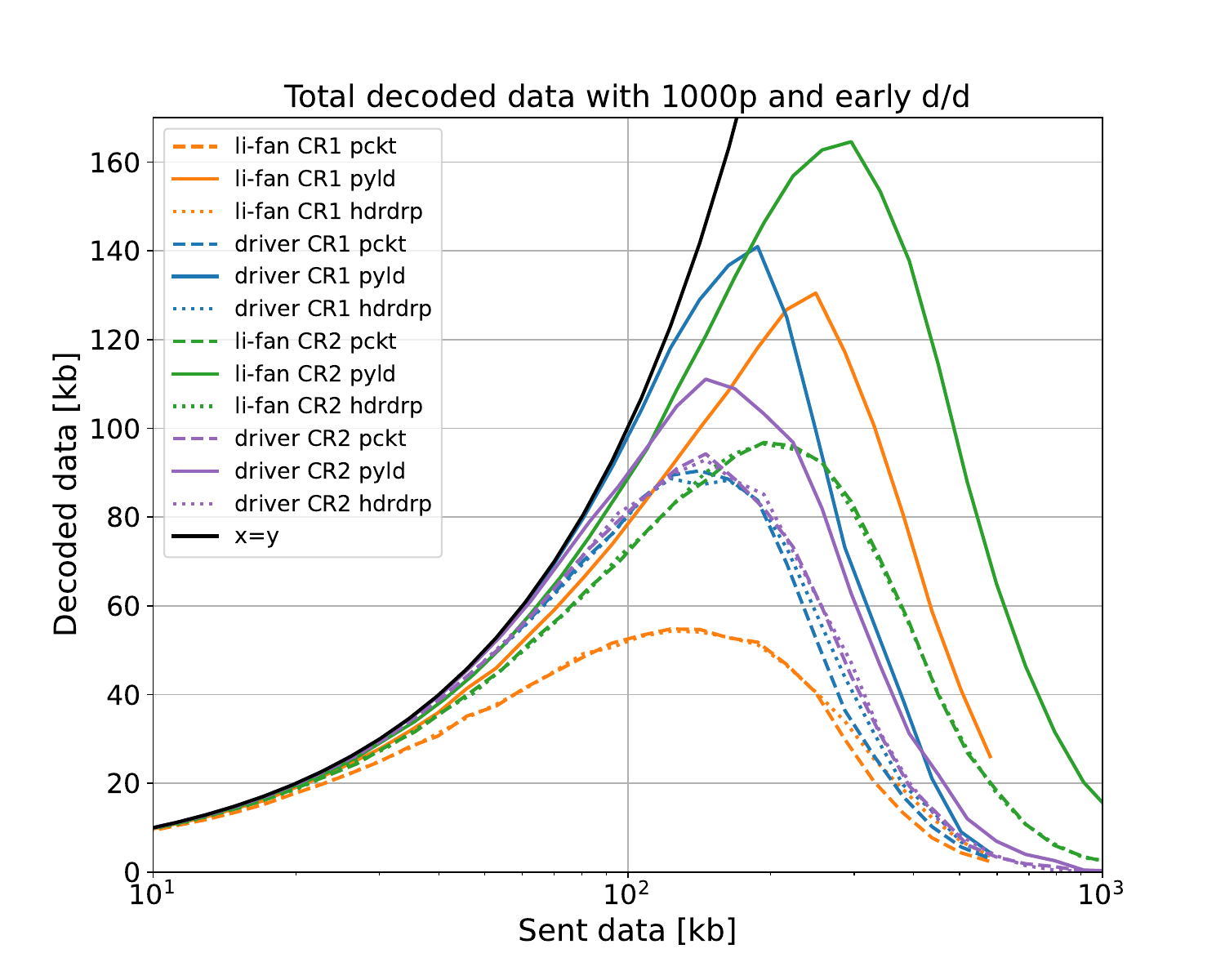}
\vspace{-0.3cm}
\label{hdrdrop1000p}}
\vspace{-0.1cm}
\caption{Total decoded data with 100 and 1000 demodulators, Early-Drop and Early-Decode.}
\label{hdrdrop}
\end{figure*}

This section delves into a comparison of packet decoding versus payload decoding. While the latter was the primary focus of the preceding section, the emphasis here shifts to packet decoding. Packet decoding implies that at least one header replica is received in addition to a successful payload decode.
For the context of this analysis, both the \textit{Early Drop} and \textit{Early Decode} mechanisms remain activated. Our metric of interest is the total decoded data compared to the total data sent in kilobytes. This approach offers a balanced comparison between CR1 and CR2. Notably, despite requiring the same number of hops, CR1 encodes less data than CR2. Thus, this metric facilitates an unbiased assessment of their performance relative to the data throughput they deliver.

\subsubsection{100 Processor Analysis}

The distinction between packet and payload decoding is brought to the fore in Fig.~\ref{hdrdrop100p}. The \textit{li-fan} and \textit{driver} FHS families are portrayed with different color codes: orange and green for \textit{li-fan}, and blue and purple for \textit{driver}. A further distinction is made between CR1 (depicted in shades of orange and blue) and CR2 (represented in shades of green and purple). Dashed lines signify packet (\textit{pckt}) decoding, while solid lines represent payload (\textit{pyld}) decoding.

The observation from the figure confirms the intuitive expectation: the rate of packet decoding, which requires both a successful payload and at least one header replica, is invariably lower than the payload-only decoding. This is depicted by the consistently higher solid lines (\textit{pyld}) compared to their dashed counterparts (\textit{pckt}) across the graph.

When considering CR1, this decrease in the packet decoding performance relative to payload decoding also manifests in the competitive dynamics between the \textit{li-fan} and \textit{driver} families. Here, the \textit{li-fan} family fails to outperform the \textit{driver} family across the entirety of the depicted range.

Turning our attention to CR2, a similar trend unfolds. Although \textit{li-fan} had previously displayed a higher efficiency, this distinction diminishes when pitted against \textit{driver} in the context of packet decoding. The relative performance advantage of \textit{li-fan} observed in the payload-only decoding scenario wanes, cementing the assertion that the header replica decoding requirement poses a challenge to both families, altering their relative efficiencies. For a dense network, \textit{li-fan} with CR2 achieves the best performance.

\subsubsection{1000 Processor Analysis}

With the enhanced processing capacity depicted in Figure \ref{hdrdrop1000p}, we observe distinct trends compared to the prior analysis with 100p.

The increased processing capability does not shield the \textit{li-fan} family from the added complexity of \textit{pckt} decoding, as the family remains noticeably more affected than when only considering \textit{pyld} decoding. For CR1, the performance gap between the two families is quite pronounced. The \textit{driver} family outpaces the \textit{li-fan} by a substantial margin, decoding up to 60\% more data. This demonstrates the \textit{driver}'s efficiency and highlights the challenges \textit{li-fan} encounters when packet decoding is in play, even with enhanced processing capability.

For CR2, the dynamics between the two families become more intricate. The \textit{driver} family retains dominance over a broader range of transmissions before \textit{li-fan} takes the lead. This extension of the \textit{driver}'s dominant zone underscores the impact of packet decoding on the \textit{li-fan}'s performance, even in the context of CR2.

Taking a step back to compare the broader impact of the coding rates, it is evident that CR2 offers superior throughput compared to CR1. This is particularly true when the available OCW channels are kept constant for both coding rates. The data, therefore, lends itself to the idea that, in terms of raw throughput and irrespective of the FHS family in question, CR2 presents a more efficient encoding choice. Again, for a dense network, \textit{li-fan} with CR2 achieves the best performance.

\subsection{Header Drop Analysis}

\begin{figure*}[!t]
\centering
\subfloat[100 demodulators]{\includegraphics[width=1 \columnwidth]{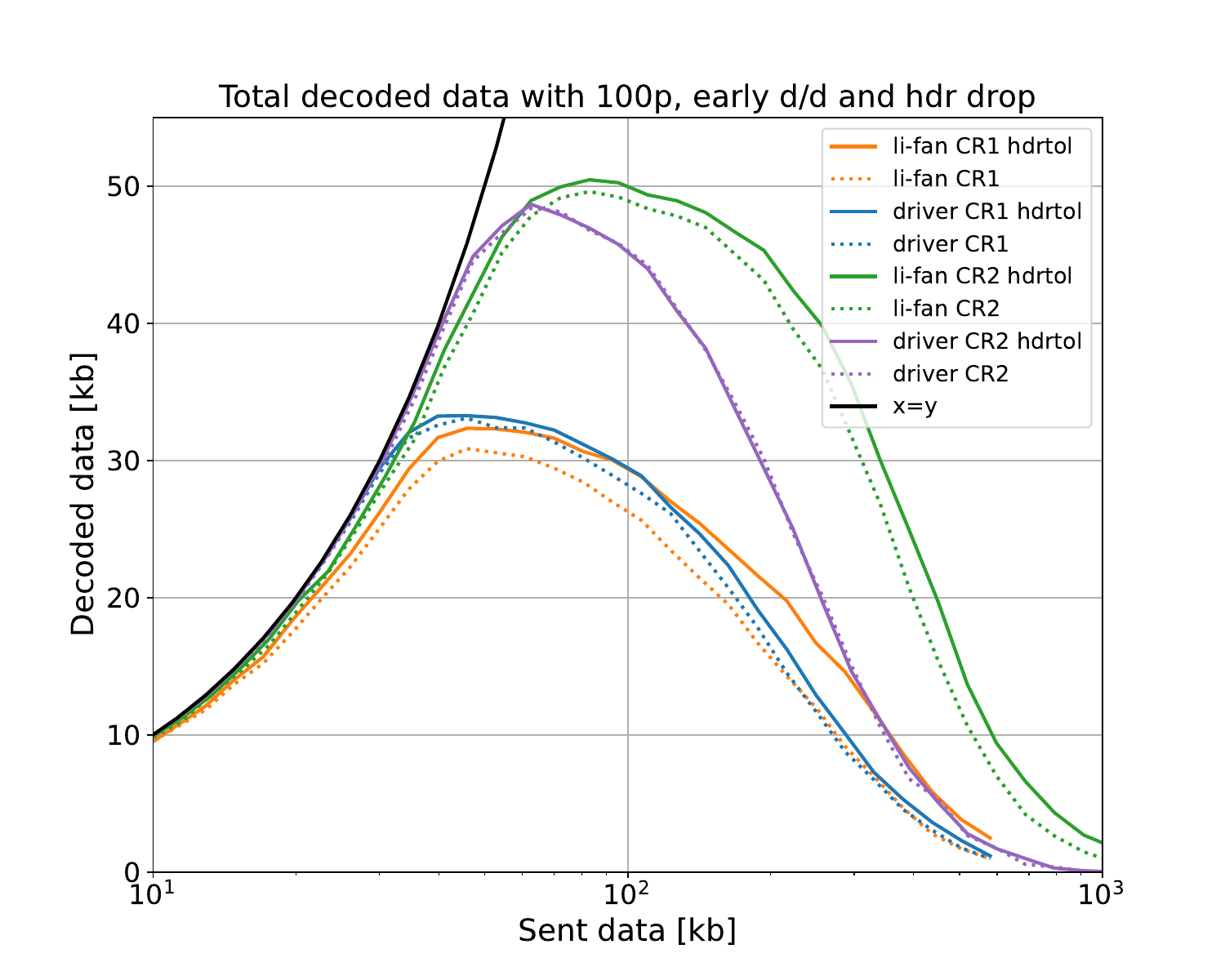}
\label{hdrtol100p}}
\hfil
\subfloat[1000 demodulators]{\includegraphics[width=1 \columnwidth]{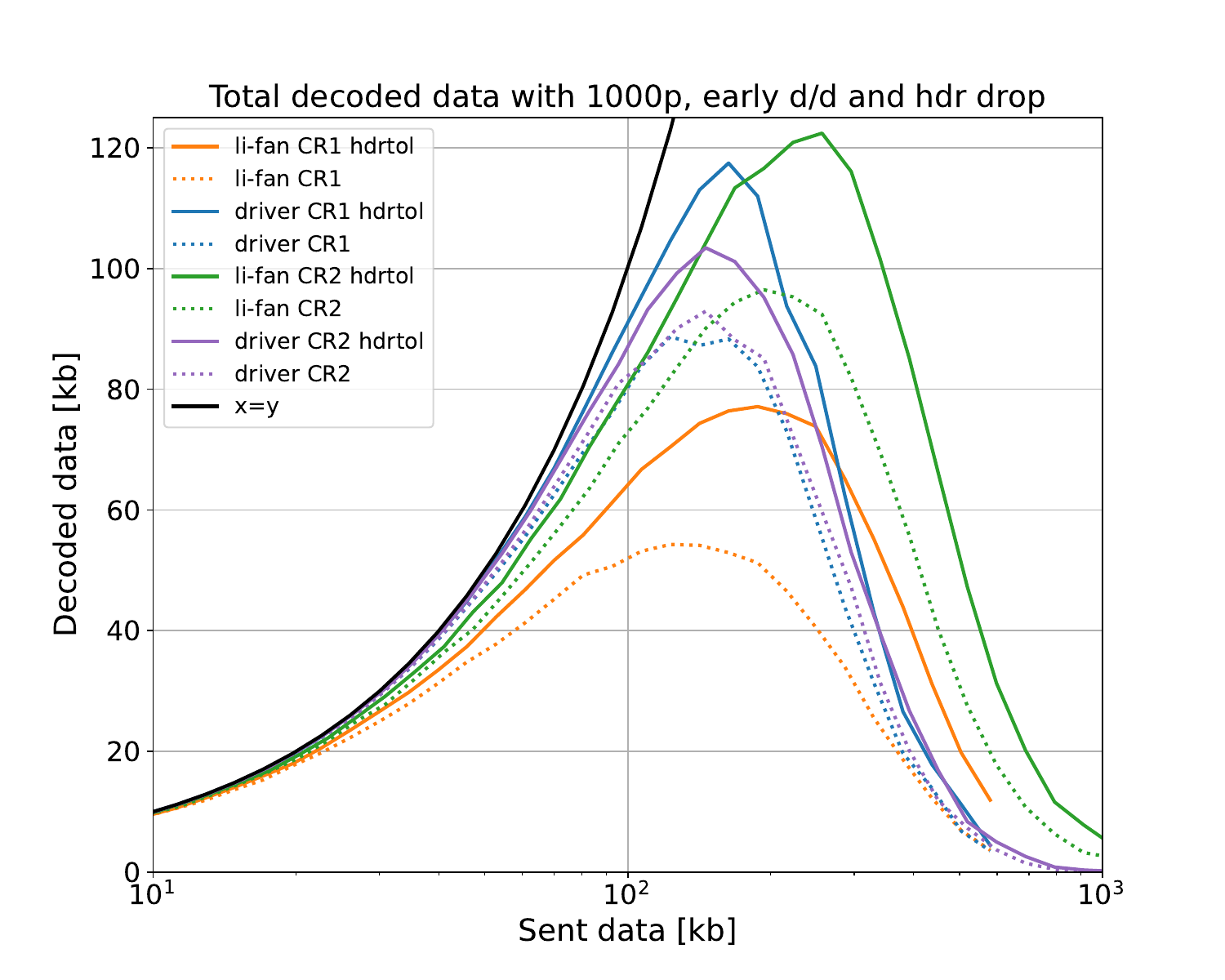}
\label{hdrtol1000p}}
\vspace{-0.2cm}
\caption{Total decoded data volume with Early-Drop, Early-Decode, Header Drop, and Header Tolerance of 4 time slots.}
\label{hdrtol}
\end{figure*}

Incorporating the \textit{Early Header Drop} mechanism brings about a significant shift in the dynamics of the decoding process, casting a spotlight on the vital role headers play in LR-FHSS networks.

In Fig.~\ref{hdrdrop100p}, which details the outcomes with 100p, a noteworthy observation is the proximity of the \textit{hdrdrp} (\textit{Early Header Drop} activated) curves to the \textit{pyld} (payload decode with \textit{Early Header Drop} off) curves. This indicates that when relieved from payloads whose headers were not successfully received, the demodulators can maintain a decoding rate almost as efficiently as if headers were non-essential. This suggests that in more constrained processing environments, the \textit{Early Header Drop} mechanism assists in focusing resources on more probable decoding successes.

However, the dynamics change as the processing power is ramped up in Fig.~\ref{hdrdrop1000p}. Here, the \textit{hdrdrp} curves converge with the \textit{pckt} curves. The vast gap between the performance of the \textit{hdrdrp} and \textit{pyld} signifies missed decoding opportunities. The divergence between these curves indicates that with more processing resources, there is substantial potential for successfully decoding payloads even when missing headers. This observation underscores the limitation of being overly reliant on headers in LR-FHSS systems.

Given the contrast observed between the \textit{hdrdrp} and \textit{pyld} results in the 1000p scenario, the findings hint at the prospective advantages of a headerless decoding mechanisms~\cite{fraire2023recovering}. If LR-FHSS networks were to develop a reliable method of decoding payloads without needing headers, it could substantially amplify the effective throughput and overall performance. With its apparent untapped potential, this direction warrants further research and could radically change LR-FHSS network methodologies.

\subsection{Header Tolerance Analysis}

This section offers an extended perspective on decoding dynamics by introducing a ``collision tolerance" for the header to properly model its more robust modulation. In Fig.~\ref{hdrtol100p} for the 100p scenario, the inclusion of the \textit{Header Tolerance} (\textit{hdrtol}) displays the immediate benefits of the flexibility offered by this model. While both FHS families witness an improvement, the \textit{li-fan} family reaps a more pronounced uptick in decoding performance. For CR2, even though the \textit{driver} family starts with a performance edge over \textit{li-fan}, this dominance is short-lived. Beyond the 60~Kb transmission threshold, \textit{li-fan} surges ahead with its decoding capacity, outperforming the \textit{driver} by as much as doubling it. Similarly, in the CR1 scenario, the earlier observations from Fig.~\ref{earlyddcr1} come to mind, where \textit{li-fan} eventually surpasses the \textit{driver} in decoding performance after a specific transmission volume.

Scaling the processing power to 1000p, as represented in Fig.~\ref{hdrtol1000p}, the benefits of the \textit{Header Tolerance} model become even more conspicuous. The resilience against header collisions boosts the decoding capacities for both families, but once again, the \textit{li-fan} stands out. The CR2 results reiterate the shift in dominance, with \textit{li-fan} pulling far ahead of the \textit{driver} family, especially after a specific transmission volume. The differential reaches a staggering 100\%, highlighting the merits of the \textit{li-fan} family in this specific scenario. The CR1 results echo a similar outcome, marking an apparent comeback for \textit{li-fan} over \textit{driver} with increasing transmission counts.

This analysis accentuates the importance of adaptability in network protocols and how slight adjustments, such as allowing for some collision tolerance in headers, can have outsized impacts on performance metrics. It underscores the potential of optimizing network parameters to adapt to specific requirements, thus enhancing overall network efficiency.

\section{Conclusions}\label{Conclusions}

The emerging integration of IoT with LEO satellites necessitates the evolution of robust, efficient, and scalable communication protocols. Our in-depth exploration into LR-FHSS modulation, mainly focusing on the design of FHS and efficient demodulator allocation strategies, has highlighted the significant promise of Wide-Gap sequences for massive LEO satellite IoT communications.
Our primary findings and contributions are summarized as follows:

\paragraph{Wide-Gap Sequence Design}
The research highlights the superior performance of the \textit{li-fan} FHS sequence family over conventional sequences within LR-FHSS contexts, notably in setups involving advanced demodulators. This insight positions the \textit{li-fan} sequences as a compelling option for enhancing LR-FHSS efficiency in massive satellite-based LR-FHSS communications.

\paragraph{Demodulator Allocation Strategies}
The introduction of ``Early-Decode" and ``Early-Drop" strategies represents a transformative approach to optimizing decoding resources for LR-FHSS gateways. These strategies proved particularly effective in boosting the performance of the \textit{li-fan} sequences, underscoring their practical benefits.

\paragraph{Combined Potential in future LR-FHSS}
By adopting a more realistic stance on collision tolerance — not immediately deeming a header as 'failed' after a single time slot collision — the integrated use of \textit{li-fan} sequences with ``Early-Decode" and ``Early-Drop" strategies pushes the scalability boundaries of LR-FHSS. This approach is especially advantageous in massive-scale environments anticipated in satellite IoT frameworks, accommodating scenarios with up to hundreds of thousands of nodes.


\begin{table}
\caption{Best FHS Family for Different LR-FHSS Configurations}
\centering
\begin{tabular}{ccccc}
\hline
\multicolumn{5}{c}{100 Decoding Processors}                                                                                                                                                                                                                                                                                                   \\ \hline
\multicolumn{1}{c|}{\multirow{2}{*}{\begin{tabular}[c]{@{}c@{}}Node\\ Count\end{tabular}}} & \multicolumn{1}{c|}{\multirow{2}{*}{\begin{tabular}[c]{@{}c@{}}Data Volume\\ CR1 in Kb\end{tabular}}} & \multicolumn{1}{c|}{}     & \multicolumn{1}{c|}{\multirow{2}{*}{\begin{tabular}[c]{@{}c@{}}Data Volume\\ CR2 in Kb\end{tabular}}} &      \\
\multicolumn{1}{c|}{}                                                                      & \multicolumn{1}{c|}{}                                                                                 & \multicolumn{1}{c|}{CR1}  & \multicolumn{1}{c|}{}                                                                                 & CR2  \\ \hline
\multicolumn{1}{c|}{0 - 500}                                                               & \multicolumn{1}{c|}{0 - 29}                                                                           & \multicolumn{1}{c|}{\textit{driver}}  & \multicolumn{1}{c|}{0 - 61}                                                                           & \textit{driver}  \\
\multicolumn{1}{c|}{500 - 1725}                                                            & \multicolumn{1}{c|}{29 - 100}                                                                         & \multicolumn{1}{c|}{\textit{driver}}  & \multicolumn{1}{c|}{61 - 209}                                                                         & \textit{li-fan} \\
\multicolumn{1}{c|}{1725 +}                                                                & \multicolumn{1}{c|}{100 +}                                                                            & \multicolumn{1}{c|}{\textit{li-fan}} & \multicolumn{1}{c|}{209 +}                                                                            & \textit{li-fan} 
\end{tabular}

\vspace{5mm}

\centering
\begin{tabular}{ccccc}
\hline
\multicolumn{5}{c}{1k Decoding Processors}                                                                                                                                                                                                                                                                                                   \\ \hline
\multicolumn{1}{c|}{\multirow{2}{*}{\begin{tabular}[c]{@{}c@{}}Node\\ Count\end{tabular}}} & \multicolumn{1}{c|}{\multirow{2}{*}{\begin{tabular}[c]{@{}c@{}}Data Volume\\ CR1 in Kb\end{tabular}}} & \multicolumn{1}{c|}{}     & \multicolumn{1}{c|}{\multirow{2}{*}{\begin{tabular}[c]{@{}c@{}}Data Volume\\ CR2 in Kb\end{tabular}}} &      \\
\multicolumn{1}{c|}{}                                                                      & \multicolumn{1}{c|}{}                                                                                 & \multicolumn{1}{c|}{CR1}  & \multicolumn{1}{c|}{}                                                                                 & CR2  \\ \hline
\multicolumn{1}{c|}{0 - 1240}                                                               & \multicolumn{1}{c|}{0 - 72}                                                                           & \multicolumn{1}{c|}{\textit{driver}}  & \multicolumn{1}{c|}{0 - 150}                                                                           & \textit{driver}  \\
\multicolumn{1}{c|}{1240 - 6550}                                                            & \multicolumn{1}{c|}{72 - 380}                                                                         & \multicolumn{1}{c|}{\textit{driver}}  & \multicolumn{1}{c|}{150 - 793}                                                                         & \textit{li-fan} \\
\multicolumn{1}{c|}{6550 +}                                                                & \multicolumn{1}{c|}{380 +}                                                                            & \multicolumn{1}{c|}{\textit{li-fan}} & \multicolumn{1}{c|}{793 +}                                                                            & \textit{li-fan} 
\end{tabular}
\label{summary}
\end{table}

Table~\ref{summary} presents a comprehensive summary of the most effective FHS families for each variant of LR-FHSS discussed in this study relative to different scales of node counts. The findings strongly suggest that the strategic integration of sequence design and resource allocation methods explored in this research pushes LR-FHSS to its scalability limits. Consequently, LR-FHSS stands out as a highly viable solution for managing the extensive IoT connectivity demands posed by the evolving direct-to-satellite IoT paradigm.

\section*{Acknowledgment}
This research has received support from the NII-INRIA MoU Grant funded by NII, the Grants-in-Aid for Scientific Research (Kakenhi 22KK0156 and 20H00592) from the Ministry
of Education, Science, Sports, and Culture of Japan, the EU's H2020 R\&D program under the Marie Skłodowska-Curie grant agreement No 101008233 (MISSION project) and the French National Research Agency (ANR) projects ANR-22-CE25-0014-01 and ANR-21-CE25-0002-01.

\bibliographystyle{IEEEtran}
\bibliography{main}

\begin{thebibliography}{10}
\providecommand{\url}[1]{#1}
\csname url@samestyle\endcsname
\providecommand{\newblock}{\relax}
\providecommand{\bibinfo}[2]{#2}
\providecommand{\BIBentrySTDinterwordspacing}{\spaceskip=0pt\relax}
\providecommand{\BIBentryALTinterwordstretchfactor}{4}
\providecommand{\BIBentryALTinterwordspacing}{\spaceskip=\fontdimen2\font plus
\BIBentryALTinterwordstretchfactor\fontdimen3\font minus
  \fontdimen4\font\relax}
\providecommand{\BIBforeignlanguage}[2]{{%
\expandafter\ifx\csname l@#1\endcsname\relax
\typeout{** WARNING: IEEEtran.bst: No hyphenation pattern has been}%
\typeout{** loaded for the language `#1'. Using the pattern for}%
\typeout{** the default language instead.}%
\else
\language=\csname l@#1\endcsname
\fi
#2}}
\providecommand{\BIBdecl}{\relax}
\BIBdecl

\bibitem{fraire2022space}
J.~A. Fraire, O.~Iova, and F.~Valois, ``Space-terrestrial integrated {I}nternet
  of {T}hings: Challenges and opportunities,'' \emph{IEEE Communications
  Magazine}, 2022.

\bibitem{de2015satellite}
M.~De~Sanctis, E.~Cianca, G.~Araniti, I.~Bisio, and R.~Prasad, ``Satellite
  communications supporting {I}nternet of remote {T}hings,'' \emph{IEEE
  Internet of Things J.}, vol.~3, no.~1, pp. 113--123, 2015.

\bibitem{fraire2019direct}
J.~A. Fraire, S.~C{\'e}spedes, and N.~Accettura, ``Direct-to-satellite {IoT}-a
  survey of the state of the art and future research perspectives: Backhauling
  the {IoT} through {LEO} satellites,'' in \emph{International Conference on
  Ad-Hoc Networks and Wireless}.\hskip 1em plus 0.5em minus 0.4em\relax
  Springer, 2019, pp. 241--258.

\bibitem{herreriaalonso23improving}
S.~Herrer\'\i{}a-Alonso, M.~Rodr\'\i{}guez-P\'erez, R.~Rodr\'\i{}guez-Rubio,
  and F.~P\'erez-Font\'an, ``Improving uplink scalability of lora-based
  direct-to-satellite iot networks,'' \emph{IEEE Internet of Things J.}, 2023.

\bibitem{alliance2018lorawan}
L.~Alliance, ``{LoRaWAN} 1.0. 3 specification,'' \emph{Technical
  Specification}, 2018.

\bibitem{semtech2020lorawan}
\BIBentryALTinterwordspacing
Semtech, ``{LoRaWAN®} protocol expands network capacity with new long range
  – frequency hopping spread spectrum technology,'' 11 2020, accessed:
  2023-10-17. [Online]. Available: \url{https://tinyurl.com/2s3ue7e7}
\BIBentrySTDinterwordspacing

\bibitem{ullah2021analysis}
M.~A. Ullah, K.~Mikhaylov, and H.~Alves, ``Analysis and simulation of {LoRaWAN}
  {LR-FHSS} for direct-to-satellite scenario,'' \emph{IEEE Wireless
  Communications Letters}, vol.~11, no.~3, pp. 548--552, 2021.

\bibitem{boquet2021lr}
G.~Boquet, P.~Tuset-Peir{\'o}, F.~Adelantado, T.~Watteyne, and X.~Vilajosana,
  ``{LR-FHSS}: Overview and performance analysis,'' \emph{IEEE Communications
  Magazine}, vol.~59, no.~3, pp. 30--36, 2021.

\bibitem{maleki2022d2d}
A.~Maleki, H.~H. Nguyen, E.~Bedeer, and R.~Barton, ``{D2D}-aided {LoRaWAN}
  {LR-FHSS} in direct-to-satellite {IoT} networks,'' \emph{arXiv preprint
  arXiv:2212.04331}, 2022.

\bibitem{maleki2022outage}
A.~Maleki, H.~H. Nguyen, and R.~Barton, ``Outage probability analysis of
  {LR-FHSS} in satellite {IoT} networks,'' \emph{IEEE Communications Letters},
  vol.~27, no.~3, pp. 946--950, 2022.

\bibitem{fraire2023recovering}
J.~A. Fraire, A.~Guitton, and O.~Iova, ``Recovering headerless frames in
  {LR-FHSS},'' \emph{arXiv preprint arXiv:2306.08360}, 2023.

\bibitem{jung2023transceiver}
S.~Jung, S.~Jeong, J.~Kang, J.~G. Ryu, and J.~Kang, ``Transceiver design and
  performance analysis for {LR-FHSS}-based direct-to-satellite {IoT},''
  \emph{arXiv preprint arXiv:2305.13779}, 2023.

\bibitem{bao2016frequency}
J.~Bao and L.~Ji, ``Frequency hopping sequences with optimal partial hamming
  correlation,'' \emph{IEEE Transactions on Information Theory}, vol.~62,
  no.~6, pp. 3768--3783, 2016.

\bibitem{ge2009optimal}
G.~Ge, Y.~Miao, and Z.~Yao, ``Optimal frequency hopping sequences: Auto-and
  cross-correlation properties,'' \emph{IEEE transactions on information
  theory}, vol.~55, no.~2, pp. 867--879, 2009.

\bibitem{peng2004lower}
D.~Peng and P.~Fan, ``Lower bounds on the hamming auto and cross correlations
  of frequency-hopping sequences,'' \emph{IEEE Transactions on Information
  Theory}, vol.~50, no.~9, pp. 2149--2154, 2004.

\bibitem{li2019new}
P.~Li, C.~Fan, Y.~Yang, and Y.~Wang, ``New bounds on wide-gap frequency-hopping
  sequences,'' \emph{IEEE Communications Letters}, vol.~23, no.~6, pp.
  1050--1053, 2019.

\bibitem{huaqing2010design}
Z.~Huaqing, ``Design and performance analysis of frequency hopping sequences
  with given minimum gap,'' in \emph{2010 International Conference on Microwave
  and Millimeter Wave Technology}.\hskip 1em plus 0.5em minus 0.4em\relax IEEE,
  2010, pp. 1271--1274.

\bibitem{bin1997one}
L.~Bin, ``One-coincidence sequences with specified distance between adjacent
  symbols for frequency-hopping multiple access,'' \emph{IEEE Transactions on
  Communications}, vol.~45, no.~4, pp. 408--410, 1997.

\bibitem{guan2014generation}
L.~Guan, Z.~Li, J.~Si, and R.~Gao, ``Generation and characteristics analysis of
  cognitive-based high-performance wide-gap {FH} sequences,'' \emph{IEEE
  Transactions on Vehicular Technology}, vol.~64, no.~11, pp. 5056--5069, 2014.

\bibitem{semtech2017sx1301}
Semtech, ``{SX1301} - digital baseband chip {LoRaWAN} macro gateways,''
  Wireless \& Sensing Products, Datasheet, 06 2017.

\bibitem{sorensen2019analysis}
R.~B. Sorensen, N.~Razmi, J.~J. Nielsen, and P.~Popovski, ``Analysis of
  {LoRaWAN} uplink with multiple demodulating paths and capture effect,'' in
  \emph{IEEE ICC (International Conference on Communications)}, 2019.

\bibitem{dalela2019lorawan}
P.~K. Dalela, S.~Sachdev, and V.~Tyagi, ``{LoRaWAN} network capacity for
  practical network planning in {I}ndia,'' in \emph{URSI AP-RASC (Asia-Pacific
  Radio Science Conference)}, 2019.

\bibitem{magrin2020study}
D.~Magrin, M.~Capuzzo, and A.~Zanella, ``A thorough study of {LoRaWAN}
  performance under different parameter settings,'' \emph{IEEE Internet of
  Things J.}, vol.~7, no.~1, pp. 116--127, 01 2020.

\bibitem{guitton2020improving}
A.~Guitton and M.~Kaneko, ``Improving {LoRa} scalability by a recursive reuse
  of demodulators,'' in \emph{GLOBECOM 2020-2020 IEEE Global Communications
  Conference}.\hskip 1em plus 0.5em minus 0.4em\relax IEEE, 2020, pp. 1--6.

\bibitem{guitton2022multi}
------, ``Multi-gateway demodulation in {LoRa},'' in \emph{GLOBECOM 2022-2022
  IEEE Global Communications Conference}.\hskip 1em plus 0.5em minus
  0.4em\relax IEEE, 2022, pp. 2008--2013.

\bibitem{applicationnote}
Semtech, ``Application note: {LR-FHSS} system performance,'' SX1261/62/LR1110,
  pp. 1--28, February 2022.

\bibitem{turletti1996gmsk}
T.~Turletti, ``{GMSK} in a nutshell,'' \emph{Telemedia Networks and Systems
  Group LCS, MIT-TR}, 1996.

\bibitem{Sornin2021-ue}
N.~Sornin, M.~Luis, T.~Eirich, T.~Kramp, and O.~Hersent, ``{RP002-1.0.3}
  {LoRaWAN} regional parameters,'' {LoRa} {Alliance} technical committee and
  others,'' LoRa Alliance, Tech. Rep., May 2021.

\bibitem{Torrieri2018}
\BIBentryALTinterwordspacing
D.~Torrieri, \emph{Frequency-Hopping Systems}.\hskip 1em plus 0.5em minus
  0.4em\relax Cham: Springer International Publishing, 2018, pp. 155--212.
  [Online]. Available: \url{https://doi.org/10.1007/978-3-319-70569-9_3}
\BIBentrySTDinterwordspacing

\bibitem{li2021constructions}
P.~Li, C.~Fan, S.~Mesnager, Y.~Yang, and Z.~Zhou, ``Constructions of optimal
  uniform wide-gap frequency-hopping sequences,'' \emph{IEEE Transactions on
  Information Theory}, vol.~68, no.~1, pp. 692--700, 2021.

\bibitem{sarwate1994reed}
D.~Sarwate, ``Reed-solomon codes and the design of sequences for
  spread-spectrum multiple-access communications,'' \emph{Reed-Solomon Codes
  and Their Applications}, 1994.

\bibitem{lempel1974families}
A.~Lempel and H.~Greenberger, ``Families of sequences with optimal
  hamming-correlation properties,'' \emph{IEEE Transactions on Information
  Theory}, vol.~20, no.~1, pp. 90--94, 1974.

\bibitem{goresky2012algebraic}
M.~Goresky and A.~Klapper, \emph{Algebraic shift register sequences}.\hskip 1em
  plus 0.5em minus 0.4em\relax Cambridge University Press, 2012.

\bibitem{vogelgesang2021uplink}
K.~Vogelgesang, J.~A. Fraire, and H.~Hermanns, ``Uplink transmission
  probability functions for {LoRa}-based direct-to-satellite {IoT}: A case
  study,'' in \emph{2021 IEEE Global Communications Conference
  (GLOBECOM)}.\hskip 1em plus 0.5em minus 0.4em\relax IEEE, 2021, pp. 01--06.

\bibitem{gonzalez2021analysis}
T.~N. Gonz{\'a}lez, J.~L. Salamanca, S.~M. S{\'a}nchez, C.~A. Meza, and
  S.~C{\'e}spedes, ``Analysis of channel models for {LoRa}-based
  direct-to-satellite {IoT} networks served by {LEO} nanosatellites,'' in
  \emph{2021 IEEE International Conference on Communications Workshops (ICC
  Workshops)}.\hskip 1em plus 0.5em minus 0.4em\relax IEEE, 2021, pp. 1--6.

\end{thebibliography}

\end{document}